\documentclass[a4paper,fleqn]{cas-sc}

\usepackage[numbers]{natbib}
\usepackage{longtable}
\usepackage{geometry}
\usepackage{appendix}
\usepackage{float}
\usepackage{lineno}
\begin{document}
\let\WriteBookmarks\relax
\def\floatpagepagefraction{1}
\def\textpagefraction{.001}
\shorttitle{Measurement of differential cross sections of $^{12}C(n,\alpha)x$}
\shortauthors{Longxiang Liu et~al.}
%\begin{frontmatter}

\title [mode = title]{Differential cross-section measurements for neutron-induced \texorpdfstring{$\alpha$}{} production reactions on carbon across neutron energy range of 6.2 to 76 MeV}                      

%\tnotetext[1]{Supported by the National Natural Science Foundation of China(Grant Nos.11905274,11875311,11775290,U2032146,11805212),the Natural Science Foundation of Inner Mongolia,China(Grant Nos.2019JQ01),the National Key Research and Development Program of China (Nos. 2016YFA0400502, 2018YFA0404404).}

\author[1]{Longxiang Liu}[style=chinese]
\cormark[1]
\ead{liulongxiang@zjlab.org.cn}
\credit{Data curation, Writing - Original draft preparation}

\address[1]{Shanghai Advanced Research Institute,Chinese Academy of Sciences (CAS),Shanghai 201210, China}
\author[2,3,4]{Kang Sun}[style=chinese]
\cormark[2]
\ead{sunkang@ihep.ac.cn}
\author[2,3]{Han Yi}[style=chinese]
\author[1]{Fei Lu}[style=chinese]
\cormark[3]
\ead{lufei@sari.ac.cn}
\author[5]{Danyang Pang}[style=chinese]
\author[1]{Hongwei Wang}[style=chinese]
\author[1]{Gongtao Fan}[style=chinese]
\author[1]{Xiguang Cao}[style=chinese]

\author[1]{Longlong Song}[style=chinese]
\author[7]{Suyalatu Zhang}[style=chinese]
\author[7]{Dexin Wang}[style=chinese]
\author[4,8]{Xinxiang Li}[style=chinese]
\author[8,9]{Yuxuan Yang}[style=chinese]
\author[4,8]{Pan Kuang}[style=chinese]
\author[4,8]{Xinrong Hu}[style=chinese]
\author[4,8]{Zirui Hao}[style=chinese]
\author[4,8]{Sheng Jin}[style=chinese]
\author[8,10]{Kaijie Chen}[style=chinese]

\author[2,3]{Wei Jiang}[style=chinese]
\author[2,3,6]{Ruirui Fan}[style=chinese]
\author[2,3,6]{Jingyu Tang}[style=chinese]
%\author[2,3]{Hantao Jing}[style=chinese]
\author[2,3,6]{Qiang Li}[style=chinese]
\author[2,3]{Yonghao Chen}[style=chinese]
\author[2,3]{Zhixin Tan}[style=chinese]
\author[2,3]{Yang Li}[style=chinese]

\author[11]{Shilun Jin}[style=chinese]

\author[12]{Yugang Ma}[style=chinese]
%\author[1]{Wenqing Shen}[style=chinese]

\address[2]{Institute of High Energy Physics, Chinese Academy of Sciences (CAS), Beijing 100049, China}

\address[3]{Spallation Neutron Source Science Center, Dongguan 523803, China}

\address[4]{University of Chinese Academy of Sciences, Beijing 100080, China}

\address[5]{School of Physics and Beijing Key Laboratory of Advanced Nuclear Materials and Physics, Beihang University, Beijing 100191, China}

\address[6]{State Key Laboratory of Particle Detection and Electronics,Institute of High Energy Physics, Chinese Academy of Sciences (CAS), Beijing 100049, China}

\address[7]{Institute of Nuclear Physics, Inner Mongolia Minzu University, Tongliao 028000, China}

\address[8]{Shanghai Institute of Applied Pysicas,Chinese Academy of Sciences (CAS),Shanghai 201800, China}

\address[9]{School of Physics and Microelectronics, Zhengzhou University, Zhengzhou 450001, China}

\address[10]{ShanghaiTech University, Shanghai 200120, China}

\address[11]{Institute of Modern Physics, Chinese Academy of Sciences (CAS), Lanzhou 730000, China}

\address[12]{Key Laboratory of Nuclear Physics and Ion-beam Application(MOE),Institute of Modern Physics,Fudan University, Shanghai 200433, China}

\cortext[cor1]{Corresponding author}
\cortext[cor2]{First Author and Second Author contribute equally to this work.}
\cortext[cor3]{Corresponding author}

\begin{abstract}
Angle-differential cross sections for neutron-induced $\alpha$ production in carbon were determined at thirty discrete neutron energy levels ranging from 6.2 to 76 MeV at the Back-n white neutron source of the China Spallation Neutron Source. 
Utilizing the ${\Delta}E-E$ telescopes within the Light-charged Particle Detector Array spanning angular measurements from $24.5^{\circ}$ to $155.5^{\circ}$ in the laboratory frame, the $^{12}C(n,\alpha)$x reaction cross sections were obtained. 
These experimental findings exhibit a strong concordance with prior results and have been benchmarked against theoretical estimates from codes such as TALYS, Geant4, and assessments from the ENDF/B-VIII.0 database. Remarkably, distinct resonance-like features were observed at neutron energies of 13.7, 22.4, 29.5, and 61.8 MeV, marking their first-time identification in the literature.
Furthermore, a comparative analysis involving the theoretical Distorted Wave Born Approximation was conducted.
\end{abstract}

\maketitle

\section{Introduction}

The study of neutron-induced reactions plays a pivotal role in enhancing our scientific comprehension and technical capabilities in several domains, including energy generation in nuclear reactors, astrophysics, and material science. In particular, the interactions between neutrons and carbon nuclei are of critical importance due to carbon's ubiquity and its diverse applications. Carbon is not only integral as a moderator and biological shielding material in nuclear reactors but also figures prominently in stellar processes that influence the nucleosynthesis of heavier elements within the interstellar medium. As such, detailed knowledge of neutron-carbon interactions is indispensable for ensuring the safe operation of nuclear installations and for advancing our understanding of astrophysical phenomena.

The significance of carbon in both terrestrial and cosmic environments cannot be overstated. Represented primarily by the nuclide $^{12}C$, it owes much of its stellar abundance to its formation through the triple-$\alpha$ process, which includes the resonant production of the $^{12}C$ Hoyle state, a phenomenon thoroughly documented in the literature~\cite{Ajzenbrg1990, PhysRev.92.649, Fynbo2005}. This state is accessible via the inelastic scattering of $^{12}C$ by neutrons and stands as a cornerstone for grasping the synthesis of carbon in stars~\cite{Bishop2022}. Despite the resonance's crucial nature, the precise structure of the $^{12}C$ Hoyle state is not fully understood, rendering it an engaging topic for the exploration of $\alpha$-clustering—a structure where carbon's nucleons arrange themselves into three $\alpha$-particle ensembles, with the Hoyle state exhibiting a heightened $\alpha$-condensed character~\cite{PhysRevLett.119.132502, PhysRevLett.113.032506, Shi2021, WANG2022137198}. Investigations have extended this concept to examine "Hoyle-like" states, wherein similar cluster phenomena occur in neighboring nuclides such as $^{13}C$, where the gaseous nature of the state manifests through the non-local movement of clusters in the potential well near the disintegration threshold~\cite{C13Kawabata2008, C13Yamada2008}.

Furthermore, carbon is being actively considered for its potential as a fusion reactor first wall material due to its capacity to withstand intense neutron fluxes. This has implications for not only fusion technology but also extends to the fabrication of advanced materials such as silicon carbide, which relies on the understanding of carbon chemistry under extreme conditions~\cite{Kondo}. For comprehensive safety assessments and material integrity analysis, the accurate computation of the double-differential cross section (DDX) of the $^{12}C(n,n^{'}+3\alpha)$ reaction, involving complex multi-particle emission channels, is integral. Although there have been concerted endeavors to measure the $\alpha$ emission resulting from the bombardment of carbon with 14 MeV neutrons, these investigations reveal that a paucity of experimental data persists across a broad spectrum of neutron energies, ultimately restricting the ability to establish robust and predictive reaction models~\cite{Kondo, Haight, Kital}.

Despite the significance of these interactions, the body of experimental data pertaining to neutron-induced alpha production from carbon, particularly at the higher end of the neutron energy spectrum, remains limited. This deficiency of experimental data translates into a lack of confidence in the simulation and predictive capabilities of current nuclear reaction models, which are otherwise instrumental in the design and analysis of nuclear technology and research.This research aims to bridge this gap by providing a meticulously acquired data set pertaining to the angle-differential cross sections of the $^{12}C(n,\alpha)x$ reaction over a neutron energy span from 6.2 to 76 MeV. The groundbreaking experiments were executed at the China Spallation Neutron Source (CSNS) – a facility known for its robust and versatile back-n white neutron source that is well-suited for broad-spectrum neutron research~\cite{ZHANG2018212,Tang2021,2021Measurements,2017Physical,Jiang2022Measurement,Hu_2022,Li_2022}.To conduct these experiments, we utilized the technologically advanced ${\Delta}E-E$ telescopes situated within the Light-charged Particle Detector Array (LPDA) system, effectively spanning angular measurements from $24.5^{\circ}$ to $155.5^{\circ}$ in the laboratory system~\cite{FAN2020164343,JIANG2020164126,Measurements2019,Bai_2020,Cui2021,Measurement2021}. Capturing the data across this wide angular range is key to enhancing our comprehension of the reaction's angular differential cross-section and advancing theoretical predictions.

Additionally, the experimental findings from this study were thoughtfully compared to theoretical models and previous experiments, as well as scrutinized alongside calculated predictions using established reaction codes such as TALYS and Geant4, and evaluated nuclear reaction data from the ENDF/B-VIII.0 library. In doing so, we aim to establish a benchmark for the accuracy of current theoretical approaches and to validate existing data libraries while providing critical experimental data to inform future developments.Most notably, this study takes advantage of the Distorted Wave Born Approximation (DWBA) to provide a theoretical perspective against which to measure our experimental data. The emergence of novel structures at certain neutron energies, as revealed by our research, offers fresh insights into the nuclear reaction process for neutrons interacting with carbon nuclei and paves the way for deeper exploration and understanding of these reactions.

\section{EXPERIMENTAL SETUP}

The CSNS facility employs a Rapid Cycling Synchrotron (RCS) to generate a proton beam, which has an energy of 1.6 GeV and a power output of 100 kW. This beam bombards a tungsten target at a frequency of 25 Hz, thereby producing spallation neutrons for experimental purposes. Measurements were carried out at Endstation 1 of the CSNS white neutron source, which delivers a broad neutron energy spectrum that extends from 0.5 eV to greater than 100 MeV through the Back-n beamline~\cite{Bai_2020}. Bending magnets are used to deflect charged particles emanating from the tungsten target to minimize their interference with experimental readings. The proton beam is structured in a double-bunch mode, wherein each bunch has a standard deviation of 25.5 ns and is separated by a 410-ns interval~\cite{Chen2019}. The neutron time-of-flight tunnel measures roughly 58 m in length from the spallation source to the sample location, where the neutron beam, being about 20 mm in diameter, achieves flux values up to $1.6\times10^{6} n/cm^{2}/s$~\cite{An_2017}.

%Fig.\ref{ctarget} shows the samples and the sample holder.

To measure the $^{12}C(n,\alpha)x$ reactions, three distinct targets were employed: a high-purity graphite foil (referred to here as the carbon target) with a thickness of 2.219 mg/cm$^2$; an empty target outfitted with a target holder analogous to the carbon target's, intended for instrumental and neutron beam background assessments; and $\alpha$ sources to evaluate detector signal fidelity. Throughout the duration of the experiment, the target's rotation angle was maintained at a constant $30^{\circ}$ relative to the direction of the incident neutron beam.

%\begin{figure}
%	\centering
%	\includegraphics[scale=.1]{figs/CTarget_Pic.jpg}
%	\caption{\centering(color online) Samples and sample holder. From top to bottom: carbon sample, empty target, empty target, and $\alpha$ sources.}
%	\label{ctarget}
%\end{figure}

\begin{figure}
	\centering
	%\includegraphics[scale=0.3]{figs/Detector_Pic.jpg}
	%\caption{\centering(color online) (a) Picture of the LPDA placed in the vacuum chamber. (b) Schematic drawing of the detector positions.}
	\includegraphics[scale=0.3]{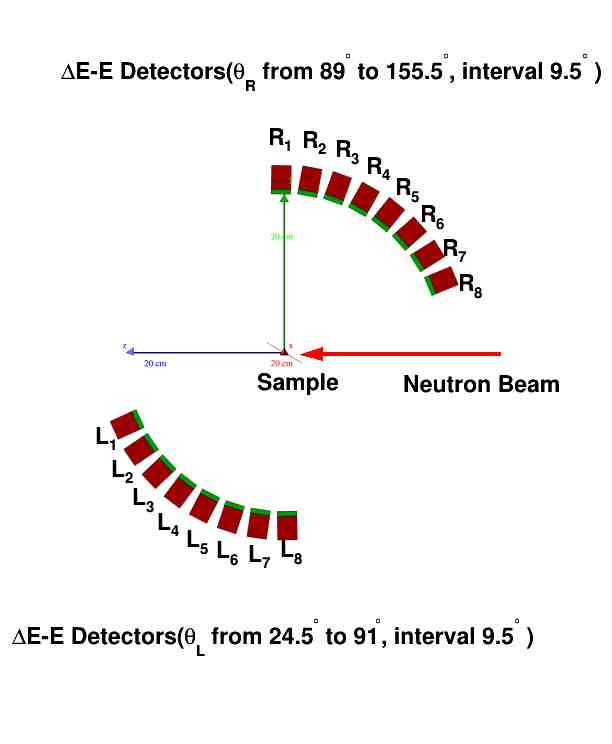}
	\caption{\centering(color online) Schematic drawing of the detector positions.}
	\label{detector_p}
\end{figure}

The cross section of the $^{12}C(n, \alpha)x$ reaction was determined using the LPDA detector system. Figure \ref{detector_p} depicts the experimental arrangement, which consisted of sixteen ${\Delta}E-E$ telescopes situated in a vacuum chamber. Each telescope was composed of a Low-Pressure Multi-Wire Proportional Chamber (LPMWPC) as the ${\Delta}E$ detector, a 300 $\mu m$-thick Si-PIN detector that served as the E or ${\Delta}E$ detector, and a 30 mm-thick CsI scintillator employed as the total E detector.

The telescopes were arrayed in the forward hemisphere, oriented at angles ranging from $24.5^{\circ}$ to $155.5^{\circ}$ relative to the incident neutron beam. The left-side telescopes, at angles ${\theta}_{L} = 24.5^{\circ}, 34.0^{\circ}, \ldots, 91.0^{\circ}$, were designated as $L_{1}, L_{2}, \ldots, L_{8}$. Conversely, the right-side sets, positioned at angles ${\theta}_{R} = 89.0^{\circ}, 98.5^{\circ}, \ldots, 155.5^{\circ}$, were denoted as $R_{1}, R_{2}, \ldots, R_{8}$.
The separation between the detectors and the target was maintained at 20.0 cm. Each detector subtended a solid angle of approximately 1.2 sr, as calculated by a Geant4 simulation. This simulation also accounted for errors attributable to the beam spot size~\cite{Geant4}.

\section{DATA ANALYSIS}

We conducted digital signal processing analysis of detector waveforms using ROOT-based programs. These programs transformed signals into Gaussian distributions and applied smoothing and bandpass filtering to ascertain event amplitude and timing. An simulated RC-CR shaping program was incorporated to zero the baseline and shape waveforms. To counteract electromagnetic noise, a simulated band-pass filter code was designed to enhance the signal-to-noise ratio by eliminating noise. The program's two-stage filtering began with the RC-CR code for baseline shaping, followed by active second-order filters to suppress extraneous frequencies and amplify target signals. Moreover, precise event timing employed a constant fraction discrimination technique, recognized for its efficacy in experimental physics.

\begin{figure}
	\centering
	\includegraphics[scale=0.8]{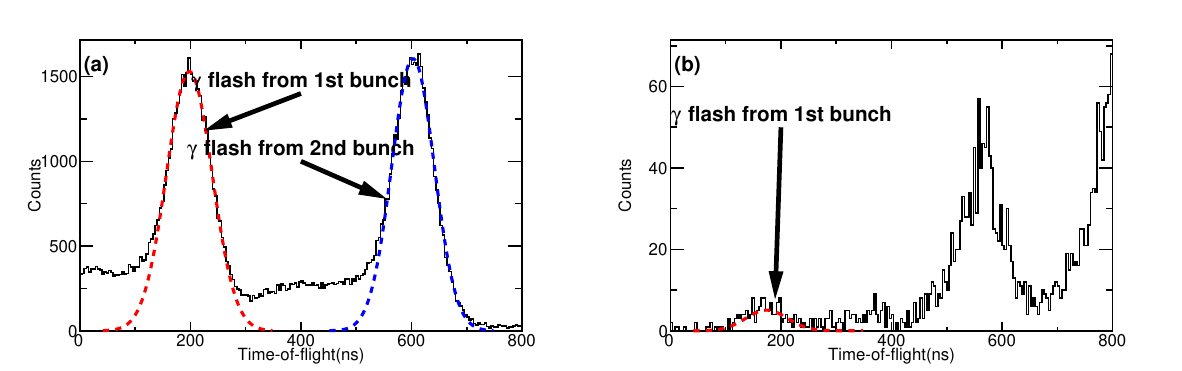}
	\caption{\centering (color online) TOF spectra of the $\gamma$ flash detected by the Si-PIN (a) and the LPMWPC (b) detectors at $\theta_{L} = 24.5^{\circ}$. The semi-transparent red and blue lines are the fitting curves corresponding to the first and second proton bunches, respectively.}
	\label{gammaflash_p}
\end{figure}

Neutron event energies are determined with the Time-of-Flight (TOF) technique, where the $\gamma$ flash from spallation, denoted as $T_{\gamma}$, establishes the peak. Figure \ref{gammaflash_p} compares the TOF $\gamma$ flash distributions for the Si-PIN and LPMWPC detectors at $\theta_{L} = 24.5^{\circ}$. The Si-PIN TOF distribution fits a double Gaussian with a 39 ns deviation attributed to the proton pulse, and the detector's time resolution is under 10 ns, compared to the data acquisition’s sub-1 ns resolution. The LPMWPC distribution shows a single peak due to its lower gamma detection efficacy. Signals within 700 ns from Si-PIN and LPMWPC are linked when calibrated with an $\alpha$ source. Timing corrections for the LPMWPC are informed by analyzing discrepancies in coincidence timing with the Si-PIN.

The formula for calculating neutron TOF is given as follows:
\begin{equation}\label{TOF_e}
	TOF=T_{\alpha}-(T_{\gamma}-\frac{L}{c})
\end{equation}
where $T_{\alpha}$ represents the arrival time of alpha events, $L$ denotes the flight length of neutrons, and $c$ signifies the velocity of light. The energy possessed by a neutron can be determined from its TOF while considering relativistic effects.

\begin{figure}
	\centering
	\includegraphics[scale=0.8]{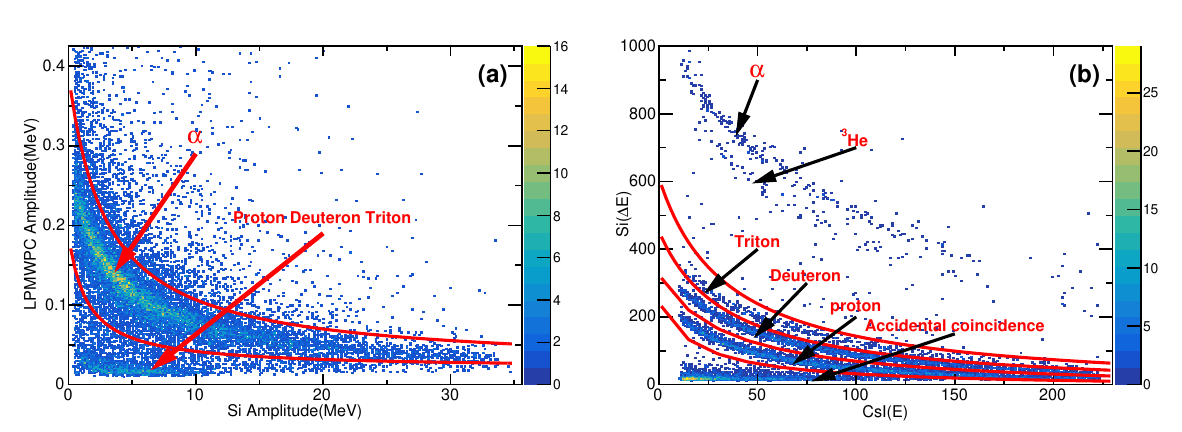}
	\caption{\centering (color online) 2-D spectra of the LPMWPC(${\Delta}E$)-Si($E$) signal amplitudes (a) and the Si(${\Delta}E$)-CsI($E$) signal amplitudes (b) at $\theta_{L} = 24.5^{\circ}$ from the carbon sample measurement.}
	\label{dEE_p}
\end{figure}

To distinguish $\alpha$ events from the carbon and the empty target, ${\Delta}E-E$ spectra were collected for all thirteen telescope arrays. Figure \ref{dEE_p} presents the two-dimensional (2-D) ${\Delta}E-E$ distributions, specifically, LPMWPC(${\Delta}E$)-Si($E$) and Si(${\Delta}E$)-CsI($E$) amplitude plots at $\theta_{L} = 24.5^{\circ}$, derived from the carbon target. The corresponding distributions for the empty target were predominantly void, confirming an absence of charged particle interference in both the neutron beam radiation and its interaction with the aluminum sample holder. During $\alpha$ particle identification, minor misclassification with $^{3}He$ nuclei occurred but was negligible and could be excluded from consideration.

\begin{figure}
	\centering
	\includegraphics[scale=0.8]{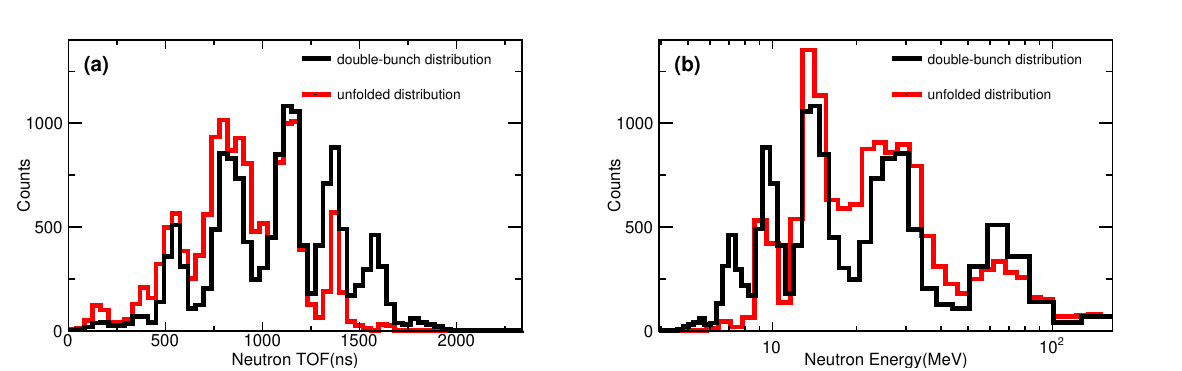}
	\caption{\centering (color online) Neutron TOF distribution (a) and neutron energy distribution (b) of double-bunch and unfolded $\alpha$ counts at $\theta_{L} = 24.5^{\circ}$.}
	\label{unfold_p}
\end{figure}

Neutrons were produced by bombarding a tungsten target with a double-bunched proton beam, spaced by 410 ns intervals. This led to pairs of potential neutron energy values, with their disparity widening at higher neutron energies. To rectify this, we applied an iterative Bayesian algorithm~\cite{Yi_2020} to deconvolute the overlapping effects induced by the double-bunched beam. 
The algorithm is defined by the following expression:
\begin{equation}\label{ufold_e}
	S_{i}^{(k+1)}=D_{i}\dfrac{S_{i}^{(k)}}{S_{i-\Delta}^{(k)}+S_{i}^{(k)}}+D_{i+\Delta}\dfrac{S_{i}^{(k)}}{S_{i}^{(k)}+S_{i+\Delta}^{(k)}}
\end{equation}
where $S_{i}$ represents the bin count in single-bunch mode, $D_{i}$ denotes the bin count measured in double-bunch mode, $\Delta$ corresponds to the number of bins associated with a 410 ns offset ($\Delta=410ns/w$, where $w$ is the bin width measured in ns), and the superscript $(k)$  indicates the $k$-th iteration.
In this study, we utilized the double-bunch unfolding software designed for the Back-n white neutron source at the China Spallation Neutron Source (CSNS). We tested algorithms employing an iterative Bayesian approach using experimental data. Figure \ref{unfold_p}(a) illustrates the Time-of-Flight (TOF) neutron distributions, and Figure \ref{unfold_p}(b) shows the energy distribution of unfolded $\alpha$-particle counts acquired with double-bunching at an angle of $\theta_{L} = 24.5^{\circ}$.

The angle differential cross-section $\sigma_{E-bin,\theta}$ is calculated by
\begin{equation}\label{cs_e}
	\sigma_{E-bin,\theta}=\frac{N^{\alpha}_{E-bin,\theta}}{\phi_{E-bin}\eta\Omega_{\theta}N_{C}}
\end{equation}
where $N^{\alpha}_{E-bin,\theta}$ represents the total net count of $\alpha$ events, $\phi_{E-bin}$ notes the neutron fluence while $\eta$ denotes detector efficiency. Furthermore, $\Omega_{\theta}$ indicates the detection solid angle corresponding to each silicon detector used in experiment, and $N_{C}$ signifies number of $^{12}C$ nuclei.
The subscripts $E-bin$ and $\theta$ refer to the neutron energy
bin and the detection angle respectively.
Table \ref{error_t} provides a list detailing sources and magnitude of uncertainties observed during this study.
Simulations suggest that $\alpha$ particles detected by detectors have a threshold value of 2.5 MeV due to carbon target thickness.
\begin{table}[width=0.9\linewidth,cols=2,pos=htb]
	\caption{ The sources  and magnitudes of uncertainties.}\label{error_t}	
	\begin{tabular*}{\tblwidth}{@{} LL@{} }
		\toprule
		Sources of the uncertainty & Magnitude (\%) \\
		\midrule
		neutron fluence ($\phi_{E-bin}$) & 2.3-4.5$^{a}$ \\
	    ${\Delta}E-E$ identification of $\alpha$ ($N^{\alpha}_{E-bin,\theta}$) & 5.0-14.6  \\
	    detection solid angle $\Omega_{\theta}$& 1.9\\
	    detector efficiency $\eta$& 3.0\\
	    Uncertainty from $N_{C}$ & 2.0\\
	    uncertainty of neutron energy $E_{n}$ (horizontal error) & 4.1-17.2$^{a}$ \\
	    unfolding of the expanding of neutron energy due to the double-bunched\\ operation mode ($N^{\alpha}_{E-bin,\theta}$)& 3.0-8.1 \\
	    Overall uncertainty & 8.5-24.7\\ 
		\bottomrule
		a:Uncertainties for 6.2 MeV $\leq$ En $\leq$ 76 MeV
	\end{tabular*}
\end{table}
\section{RESULTS AND DISCUSSION}
\begin{figure}
	\centering
	\includegraphics[scale=0.7]{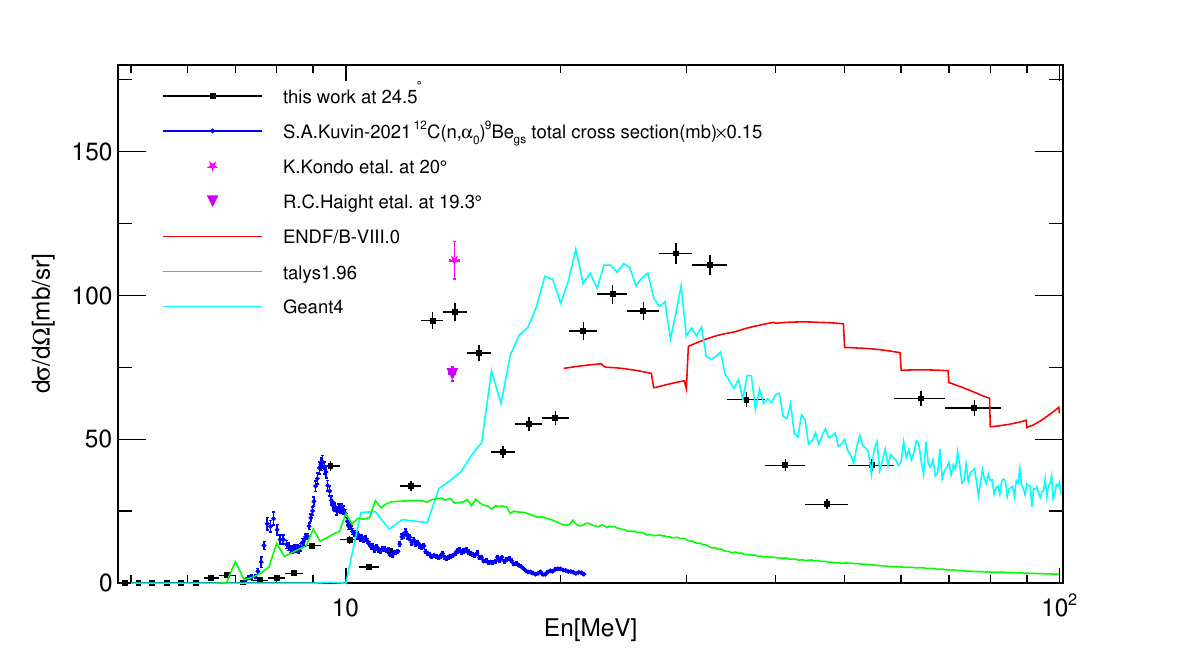}
	\caption{\centering (color online) Differential cross sections of the $^{12}C(n,\alpha)x$ reactions measured at $\theta_{L} = 24.5^{\circ}$ compared with results from previous measurements, the evaluated data, and the simulation.(Only statistical uncertainties were included)}
	\label{aCS_p}
\end{figure}

The differential cross sections of the $^{12}C(n,\alpha)x$ reactions were measured at various angles in the laboratory system, including $\theta_{L} = 24.5^{\circ}, 34.0^{\circ}, 43.5^{\circ}, 53.0^{\circ}, 62.5^{\circ}, 72.0^{\circ}, 81.5 ^{\circ}, 91.0^{\circ}$ and $\theta_{R} = 98.5^{\circ}, 108.0^{\circ}, 127.0^{\circ}, 136.5^{\circ}$, and $146.0 ^{\circ}$ as shown in Fig.\ref{aCS_p}, \ref{aMACS1_p} and \ref{aMACS2_p} with tabulated data presented in Table \ref{macs_t} and \ref{macs2_t}.
The experiment was simulated using Geant4 with QBBC physics model while considering particle threshold values for each figure displayed herein.
Additionally, Talys-1.96 code was employed to calculate data using default parameters while taking into account particle thresholds; results are also presented in each figure~\cite{KONING20122841}.
ENDF/B-VIII.0 library data calculated based on particle thresholds within energy range of 20 MeV to 100 MeV~\cite{BROWN20181} are included as well.
These experimental findings suggest that modifications may be required within existing theoretical frameworks to achieve more accurate representations of physical phenomena under investigation in future studies. 
Consequently, it is essential for researchers to refine their methodologies and theoretical models by incorporating the experimental observations obtained in this work.

Furthermore, prior measurement results have been provided for comparison purposes.
In Fig.\ref{aCS_p}~\cite{PhysRevC.104.014603}, the total cross sections of the $^{12}C(n,\alpha_{0})^{9}Be_{gs}$ multiplied by a factor of 0.15 is depicted.
However, in this study,the absence of the peak near 8 MeV can be attributed to the fact that the energy of $\alpha$ particles produced by neutrons with this energy is insufficient, measuring less than 2.5 MeV.
In the prior investigations conducted by K.Kondo ~\cite{Kondo} and R.C.Haight ~\cite{Haight}, double differential cross-section measurements were obtained for $\alpha$ production reactions on carbon. These measurements were subsequently energy-integrated to derive angle-differential cross sections for comparative purposes, taking into account the $\alpha$ threshold. 
The experimental data acquired at $20^{\circ}$ exhibit a strong correlation with the findings of this study at $24.5^{\circ}$.
Contrary to theoretical predictions, peak structures were identified in this research at neutron energies of 13.7 MeV, 22.4 MeV, 29.5 MeV, and 61.8 MeV - a discovery that has not been previously reported in literature. This observation suggests distinct excited states of $^{9}Be$ as well as differing $\alpha$ emission mechanisms within $^{13}C$.
Due to statistical limitations encountered during unfolding procedures involving doubly-bunche $\alpha$ distributions, energy information for $\alpha$ particles was lost. Consequently, determining the Q-value for this reaction has proven to be a challenging task. Further research is necessary to thoroughly examine these peak structures.

\begin{figure}
	\centering
	\includegraphics[scale=0.8]{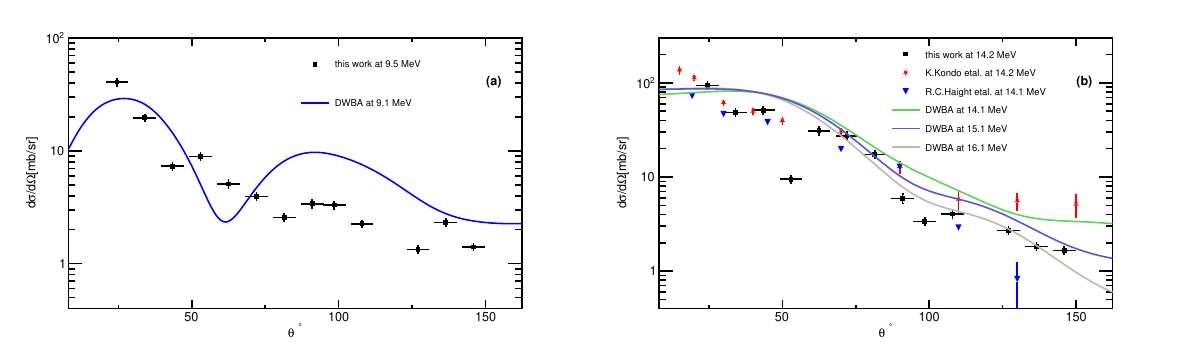}
	\caption{\centering (color online) The angle-differential cross sections of the $^{12}C(n,\alpha)x$ reactions measured at 9.5 MeV (a) and 14.2 MeV (b) compared with results from previous measurements.(Only statistical uncertainties were included)}
	\label{aACS_p}
\end{figure}
The angle-differential cross sections of $^{12}C(n,\alpha)x$ reactions measured at 14.2 MeV are depicted in Fig.\ref{aACS_p} and compared with the energy-integrated results obtained by K.Kondo and R.C.Haight. Within smaller $\theta$ ranges, the current data exhibit general agreement with previous measurements from both researchers.
However, when considering larger $\theta$ values, inconsistencies in data quality can primarily be attributed to insufficient statistics and excessive uncertainty levels. 
%The outcomes derived from our study are markedly lower than those predicted by Geant4 simulations. 

The angle-differential cross sections for the  $^{12}C(n,\alpha)^{9}Be$ reaction were computed using the Distorted Wave Born Approximation (DWBA) at incident neutron energies of 9.5 and 14.2 MeV. Optical Model Potentials (OMPs) for the entrance channel—the interaction of neutrons with the $^{12}C$ target—were derived from the Koning and Delaroche systematics~\cite{KONING2003231}. The exit channel OMPs, corresponding to the ( $\alpha+^{9}Be$ ) system, were sourced from the Perey and Perey data compilation~\cite{PEREY19761}.

In the calculations, only direct reaction mechanisms are considered. For the $^{12}C(n,\alpha)^{9}Be$ reaction, two direct processes are possible: 1) the transfer of a $^{3}He$ cluster, where the incident neutron captures a $^{3}He$ cluster from the $^{12}C$ target's ground state, resulting in an alpha particle and a residual $^{9}Be$ nucleus, and 2) the transfer of an $^{8}Be$ cluster, whereby the neutron seizes an $^{8}Be$ cluster from the $^{12}C$ ground state, forming a $^{9}Be$ nucleus and leaving behind an alpha particle as the residue. Experimentally, these processes cannot be distinguished. The theoretical amplitudes of these mechanisms are coherently summed to obtain the final cross sections. The spectroscopic amplitudes for the $^{3}He$ cluster in the ground states of the alpha particle and $^{12}C$, along with those for the $^{8}Be$ cluster in the ground states of $^{9}Be$ and $^{12}C$, are assumed to be 1.41, -1.74, 0.46, and 0.557, respectively, as adopted from reference ~\cite{KARIEM2006,KURATH1975269,JENNY2007,SINHA1930} respectively. Standard Woods-Saxon potentials are employed to compute the cluster single particle wave functions. The radius and diffuseness parameters, ${r_0, a_0}$ (expressed in femtometers), are set at {1.524, 0.5} for $^{4}He \rightarrow ^{3}He+n$~\cite{DEVRIES1972424}, {1.693, 0.65} for $^{12}C \rightarrow ^{9}Be+^{3}He$~\cite{ESWARAN1990}, and the conventional {1.25, 0.65} for $ ^{9}Be \rightarrow ^{8}Be+n$ and $^{12}C \rightarrow ^{8}Be+^{4}He$.

The calculations indicate that the transfer process of the $^{3}He$ cluster predominates at both incident energy levels. In contrast, the contributions stemming from the $^{8}Be$ transfer are negligible.

\section{CONCLUSIONS}
In this study, we present differential cross-section measurements for the $^{12}C(n,\alpha)x$ reactions at thirteen distinct scattering angles, obtained using neutrons with energies ranging from 6.2 MeV to 76 MeV at the CSNS back-n white neutron source. 
Furthermore, we include calculations by employing the Talys-1.96 code with default parameters and the Geant4 simulation framework with the QBBC physics model. 
We validated our results by comparing them with both experimental and evaluated data from previous studies and the ENDF/B-VIII.0 library. 
The observed trends and values in our data are consistent with prior research and extend the available energy range for differential cross-section measurements. 
Notably, we identified features related to neutron energies of 13.7 MeV, 22.4 MeV, 29.5 MeV, and 61.8 MeV that correspond to previously undetected excited states of $^{9}Be$ and novel $\alpha$ emission mechanisms in $^{13}C$. 
Theoretical calculations using DWBA suggest the dominance of the $^{3}He$ cluster transfer process at 9.1 and 14.1 MeV incident energies. 
These findings greatly enhance our understanding of nuclear reactions involving carbon isotopes at various energy levels and have important implications for fields such as astrophysics and nuclear engineering.

\section*{ACKNOWLEDGEMENTS}

This work is supported by the National Key Research and Development Program of China (Nos. 2023YFA1606701, 2016YFA0400502), the National Natural Science Foundation of China (Grant Nos.11905274, 11875311, 11775290, U2032146, 11805212), and the Natural Science Foundation of Inner Mongolia,China (Grant Nos.2019JQ01),
The authors are grateful to the operating crew of the CSNS white neutron source. Senior Engineer Jiming Wang, Yongli Jin and Xi Tao from the China Institute of Atomic Energy is appreciated for the support in data reading from the ENDF-B/VIII.0 library.


%\bibliographystyle{cas-model2-names}
\bibliographystyle{elsarticle-num}
%\bibliographystyle{unsrt}


\bibliography{cas-refs}

\begin{thebibliography}{10}
\expandafter\ifx\csname url\endcsname\relax
  \def\url#1{\texttt{#1}}\fi
\expandafter\ifx\csname urlprefix\endcsname\relax\def\urlprefix{URL }\fi
\expandafter\ifx\csname href\endcsname\relax
  \def\href#1#2{#2} \def\path#1{#1}\fi

\bibitem{Ajzenbrg1990}
F.~Ajzenberg-Selove,
  \href{https://www.sciencedirect.com/science/article/pii/037594749090271M}{Energy
  levels of light nuclei a = 11-12}, Nuclear Physics A 506~(1) (1990) 1--158.
\newblock \href
  {http://dx.doi.org/https://doi.org/10.1016/0375-9474(90)90271-M}
  {\path{doi:https://doi.org/10.1016/0375-9474(90)90271-M}}.
\newline\urlprefix\url{https://www.sciencedirect.com/science/article/pii/037594749090271M}

\bibitem{PhysRev.92.649}
D.~N.~F. Dunbar, R.~E. Pixley, W.~A. Wenzel, W.~Whaling,
  \href{https://link.aps.org/doi/10.1103/PhysRev.92.649}{The 7.68-mev state in
  c12}, Phys. Rev. 92 (1953) 649--650.
\newblock \href {http://dx.doi.org/10.1103/PhysRev.92.649}
  {\path{doi:10.1103/PhysRev.92.649}}.
\newline\urlprefix\url{https://link.aps.org/doi/10.1103/PhysRev.92.649}

\bibitem{Fynbo2005}
H.~O.~U. Fynbo, C.~A. Diget, U.~C. Bergmann, M.~J.~G. Borge, J.~Cederkäll,
  P.~Dendooven, L.~M. Fraile, S.~Franchoo, V.~N. Fedosseev, B.~R. Fulton,
  W.~Huang, J.~Huikari, H.~B. Jeppesen, A.~S. Jokinen, P.~Jones, B.~Jonson,
  U.~Köster, K.~Langanke, M.~Meister, T.~Nilsson, G.~Nyman, Y.~Prezado,
  K.~Riisager, S.~Rinta-Antila, O.~Tengblad, M.~Turrion, Y.~Wang, L.~Weissman,
  K.~Wilhelmsen, J.~Äystö, {The ISOLDE Collaboration},
  \href{https://doi.org/10.1038/nature03219}{Revised rates for the stellar
  triple-alpha process from measurement of c12 nuclear resonances}, Nature 433
  (2005) 136--139.
\newblock \href {http://dx.doi.org/10.1038/nature03219}
  {\path{doi:10.1038/nature03219}}.
\newline\urlprefix\url{https://doi.org/10.1038/nature03219}

\bibitem{Bishop2022}
J.~Bishop, C.~E. Parker, G.~V. Rogachev, S.~Ahn, E.~Koshchiy, K.~Brandenburg,
  C.~R. Brune, R.~J. Charity, J.~Derkin, N.~Dronchi, G.~Hamad,
  Y.~{Jones-Alberty}, T.~Kokalova, T.~N. Massey, Z.~Meisel, E.~V. Ohstrom,
  S.~N. Paneru, E.~C. Pollacco, M.~Saxena, N.~Singh, R.~Smith, L.~G. Sobotka,
  D.~Soltesz, S.~K. Subedi, A.~V. Voinov, J.~Warren, C.~Wheldon,
  \href{https://doi.org/10.1038/s41467-022-29848-7}{Neutron-upscattering
  enhancement of the triple-alpha process}, Nature Communications 13 (2022)
  1--5.
\newblock \href {http://dx.doi.org/10.1038/s41467-022-29848-7}
  {\path{doi:10.1038/s41467-022-29848-7}}.
\newline\urlprefix\url{https://doi.org/10.1038/s41467-022-29848-7}

\bibitem{PhysRevLett.119.132502}
R.~Smith, T.~Kokalova, C.~Wheldon, J.~E. Bishop, M.~Freer, N.~Curtis, D.~J.
  Parker, \href{https://link.aps.org/doi/10.1103/PhysRevLett.119.132502}{New
  measurement of the direct 3 alpha decay from the c12 hoyle state}, Phys. Rev.
  Lett. 119 (2017) 132502.
\newblock \href {http://dx.doi.org/10.1103/PhysRevLett.119.132502}
  {\path{doi:10.1103/PhysRevLett.119.132502}}.
\newline\urlprefix\url{https://link.aps.org/doi/10.1103/PhysRevLett.119.132502}

\bibitem{PhysRevLett.113.032506}
W.~B. He, Y.~G. Ma, X.~G. Cao, X.~Z. Cai, G.~Q. Zhang,
  \href{https://link.aps.org/doi/10.1103/PhysRevLett.113.032506}{Giant dipole
  resonance as a fingerprint of alpha clustering configurations in 12c and
  16o}, Phys. Rev. Lett. 113 (2014) 032506.
\newblock \href {http://dx.doi.org/10.1103/PhysRevLett.113.032506}
  {\path{doi:10.1103/PhysRevLett.113.032506}}.
\newline\urlprefix\url{https://link.aps.org/doi/10.1103/PhysRevLett.113.032506}

\bibitem{Shi2021}
C.~Z. Shi, Y.~G. Ma,
  \href{https://doi.org/10.1007/s41365-021-00897-9}{Alpha-clustering effect on
  flows of direct photons in heavy-ion collisions}, Nuclear Science and
  Techniques 32.
\newblock \href {http://dx.doi.org/10.1007/s41365-021-00897-9}
  {\path{doi:10.1007/s41365-021-00897-9}}.
\newline\urlprefix\url{https://doi.org/10.1007/s41365-021-00897-9}

\bibitem{WANG2022137198}
Y.~Z. Wang, S.~Zhang, Y.~G. Ma,
  \href{https://www.sciencedirect.com/science/article/pii/S037026932200332X}{System
  dependence of away-side broadening and alpha-clustering light nuclei
  structure effect in dihadron azimuthal correlations}, Physics Letters B 831
  (2022) 137198.
\newblock \href
  {http://dx.doi.org/https://doi.org/10.1016/j.physletb.2022.137198}
  {\path{doi:https://doi.org/10.1016/j.physletb.2022.137198}}.
\newline\urlprefix\url{https://www.sciencedirect.com/science/article/pii/S037026932200332X}

\bibitem{C13Kawabata2008}
T.~KAWABATA, Y.~SASAMOTO, Y.~MAEDA, S.~SAKAGUCHI, Y.~SHIMIZU, K.~SUDA,
  T.~UESAKA, M.~FUJIWARA, H.~HASHIMOTO, K.~HATANAKA, K.~KAWASE, H.~MATSUBARA,
  K.~NAKANISHI, Y.~TAMESHIGE, A.~TAMII, K.~ITOH, M.~ITOH, H.~P. YOSHIDA,
  Y.~KANADA-EN'YO, M.~UCHIDA,
  \href{https://doi.org/10.1142/S0218301308011112}{Cluster states in 13c and
  11b}, International Journal of Modern Physics E 17~(10) (2008) 2071--2075.
\newblock \href
  {http://arxiv.org/abs/https://doi.org/10.1142/S0218301308011112}
  {\path{arXiv:https://doi.org/10.1142/S0218301308011112}}, \href
  {http://dx.doi.org/10.1142/S0218301308011112}
  {\path{doi:10.1142/S0218301308011112}}.
\newline\urlprefix\url{https://doi.org/10.1142/S0218301308011112}

\bibitem{C13Yamada2008}
T.~YAMADA, Y.~FUNAKI, \href{https://doi.org/10.1142/S0218301308011161}{Cluster
  states and alpha particle condensation in 13c}, International Journal of
  Modern Physics E 17~(10) (2008) 2101--2105.
\newblock \href
  {http://arxiv.org/abs/https://doi.org/10.1142/S0218301308011161}
  {\path{arXiv:https://doi.org/10.1142/S0218301308011161}}, \href
  {http://dx.doi.org/10.1142/S0218301308011161}
  {\path{doi:10.1142/S0218301308011161}}.
\newline\urlprefix\url{https://doi.org/10.1142/S0218301308011161}

\bibitem{Kondo}
K.~KONDO, I.~MURATA, K.~OCHIAI, N.~KUBOTA, H.~MIYAMARU, C.~KONNO, T.~NISHITANI,
  \href{https://www.tandfonline.com/doi/abs/10.1080/18811248.2008.9711420}{Measurement
  and analysis of neutron-induced alpha particle emission double-differential
  cross section of carbon at 14.2 mev}, Journal of Nuclear Science and
  Technology 45~(2) (2008) 103--115.
\newblock \href
  {http://arxiv.org/abs/https://www.tandfonline.com/doi/pdf/10.1080/18811248.2008.9711420}
  {\path{arXiv:https://www.tandfonline.com/doi/pdf/10.1080/18811248.2008.9711420}},
  \href {http://dx.doi.org/10.1080/18811248.2008.9711420}
  {\path{doi:10.1080/18811248.2008.9711420}}.
\newline\urlprefix\url{https://www.tandfonline.com/doi/abs/10.1080/18811248.2008.9711420}

\bibitem{Haight}
R.~C. Haight, S.~M. Grimes, R.~G. Johnson, H.~H. Barschal,
  \href{https://doi.org/10.13182/NSE84-A17444}{The 12c(n,a) reaction and the
  kerma factor for carbon at en = 14.1 mev}, Nuclear Science and Engineering
  87~(1) (1984) 41--47.
\newblock \href {http://arxiv.org/abs/https://doi.org/10.13182/NSE84-A17444}
  {\path{arXiv:https://doi.org/10.13182/NSE84-A17444}}, \href
  {http://dx.doi.org/10.13182/NSE84-A17444} {\path{doi:10.13182/NSE84-A17444}}.
\newline\urlprefix\url{https://doi.org/10.13182/NSE84-A17444}

\bibitem{Kital}
R.~A. Al-Kital, R.~A. Peck, C12(n, a)be9 reaction induced by 14-mev neutrons,
  Physical Review 130~(4) (1963) 1500--1503.
\newblock \href {http://dx.doi.org/10.1103/PhysRev.130.1500}
  {\path{doi:10.1103/PhysRev.130.1500}}.

\bibitem{ZHANG2018212}
L.~Zhang, H.~Jing, J.~Tang, Q.~Li, X.~Ruan, J.~Ren, C.~Ning, Y.~Yu, Z.~Tan,
  P.~Wang, Y.~He, X.~Wang,
  \href{https://www.sciencedirect.com/science/article/pii/S0969804317309119}{Design
  of back-streaming white neutron beam line at csns}, Applied Radiation and
  Isotopes 132 (2018) 212--221.
\newblock \href
  {http://dx.doi.org/https://doi.org/10.1016/j.apradiso.2017.11.013}
  {\path{doi:https://doi.org/10.1016/j.apradiso.2017.11.013}}.
\newline\urlprefix\url{https://www.sciencedirect.com/science/article/pii/S0969804317309119}

\bibitem{Tang2021}
J.~Y. Tang, Q.~An, J.~B. Bai, J.~Bao, Y.~Bao, P.~Cao, H.~L. Chen, Q.~P. Chen,
  Y.~H. Chen, Z.~Chen, Z.~Q. Cui, R.~R. Fan, C.~Q. Feng, K.~Q. Gao, X.~L. Gao,
  M.~H. Gu, C.~C. Han, Z.~J. Han, G.~Z. He, Y.~C. He, Y.~Hong, Y.~W. Hu, H.~X.
  Huang, X.~R. Huang, H.~Y. Jiang, W.~Jiang, Z.~J. Jiang, H.~T. Jing, L.~Kang,
  B.~Li, C.~Li, J.~W. Li, Q.~Li, X.~Li, Y.~Li, J.~Liu, R.~Liu, S.~B. Liu, X.~Y.
  Liu, Z.~Long, G.~Y. Luan, C.~J. Ning, M.~C. Niu, B.~B. Qi, J.~Ren, Z.~Z. Ren,
  X.~C. Ruan, Z.~H. Song, K.~Sun, Z.~J. Sun, Z.~X. Tan, X.~Y. Tang, B.~B. Tian,
  L.~J. Wang, P.~C. Wang, Z.~H. Wang, Z.~W. Wen, X.~G. Wu, X.~Wu, L.~K. Xie,
  X.~Y. Yang, Y.~W. Yang, H.~Yi, L.~Yu, T.~Yu, Y.~J. Yu, G.~H. Zhang, L.~H.
  Zhang, Q.~W. Zhang, X.~P. Zhang, Y.~L. Zhang, Z.~Y. Zhang, L.~P. Zhou, Z.~H.
  Zhou, K.~J. Zhu, {The CSNS Back-n Collaboration}, Back-n white neutron source
  at csns and its applications, Nuclear Science and Techniques 31~(1).
\newblock \href {http://dx.doi.org/10.1007/s41365-021-00846-6}
  {\path{doi:10.1007/s41365-021-00846-6}}.

\bibitem{2021Measurements}
X.~R. Hu, G.~T. Fan, W.~Jiang, J.~Ren, L.~X. Liu, H.~W. Wang, Y.~D. Liu, X.~X.
  Li, Y.~Zhang, Z.~R. Hao,
  \href{https://doi.org/10.1007/s41365-021-00931-w}{Measurements ofthe 197au(n,
  g) crosssectionupto100kev at the csnsback-nfacility}, Nuclear Science and
  Techniques 32~(9) (2021) 10.
\newblock \href {http://dx.doi.org/10.1007/s41365-021-00931-w}
  {\path{doi:10.1007/s41365-021-00931-w}}.
\newline\urlprefix\url{https://doi.org/10.1007/s41365-021-00931-w}

\bibitem{2017Physical}
G.~Luan, Q.~Wang, J.~Bao, X.~Ruan, J.~Ren, H.~Jing, K.~Zhang, H.~Huang,
  Physical design of fast fission chamber for neutron spectrum and flux
  measurement at the csns backstreaming white neutron source, He Jishu/Nuclear
  Techniques 40.
\newblock \href {http://dx.doi.org/10.11889/j.0253-3219.2017.hjs.40.110501}
  {\path{doi:10.11889/j.0253-3219.2017.hjs.40.110501}}.

\bibitem{Jiang2022Measurement}
B.~Jiang, J.~Han, J.~Ren, W.~Jiang, X.~Wang, Z.~Guo, J.~Zhang, J.~Hu, J.~Chen,
  X.~Cai, H.~Wang, L.~Liu, X.~Li, X.~Hu, Y.~Zhang,
  \href{https://doi.org/10.1088/1674-1056/ac5394}{Measurement of 232th (n, g)
  cross section at the csns back-n facility in the unresolved resonance region
  from 4 kev to 100 kev}, Chinese Physics B 31~(6) (2022) 060101.
\newblock \href {http://dx.doi.org/10.1088/1674-1056/ac5394}
  {\path{doi:10.1088/1674-1056/ac5394}}.
\newline\urlprefix\url{https://doi.org/10.1088/1674-1056/ac5394}

\bibitem{Hu_2022}
X.~R. Hu, L.~X. Liu, W.~Jiang, J.~Ren, G.~T. Fan, H.~W. Wang, X.~G. Cao, L.~L.
  Song, Y.~D. Liu, Y.~Zhang, X.~X. Li, Z.~R. Hao, P.~Kuang, X.~H. Wang, J.~F.
  Hu, B.~Jiang, D.~X. Wang, S.~Zhang, Z.~D. An, Y.~T. Wang, C.~W. Ma, J.~J. He,
  J.~Su, L.~Y. Zhang, Y.~X. Yang, S.~Jin, K.~J. Chen,
  \href{https://doi.org/10.1088/1674-1056/ac6ee2}{New experimental measurement
  of natse(n, g) cross section between 1 ev to 1 kev at the csns back-n
  facility}, Chinese Physics B 31~(8) (2022) 080101.
\newblock \href {http://dx.doi.org/10.1088/1674-1056/ac6ee2}
  {\path{doi:10.1088/1674-1056/ac6ee2}}.
\newline\urlprefix\url{https://doi.org/10.1088/1674-1056/ac6ee2}

\bibitem{Li_2022}
X.~X. Li, L.~X. Liu, W.~Jiang, J.~Ren, H.~W. Wang, G.~T. Fan, J.~J. He, X.~G.
  Cao, L.~L. Song, Y.~Zhang, X.~R. Hu, Z.~R. Hao, P.~Kuang, B.~Jiang, X.~H.
  Wang, J.~F. Hu, J.~C. Wang, D.~X. Wang, S.~Y. Zhang, Y.~D. Liu, X.~Ma, C.~W.
  Ma, Y.~T. Wang, Z.~D. An, J.~Su, L.~Y. Zhang, Y.~X. Yang, W.~B. Liu, W.~Q.
  Su, S.~Jin, K.~J. Chen,
  \href{https://doi.org/10.1088/1674-1056/ac48fd}{Measurements of the 107ag
  neutron capture cross sections with pulse height weighting technique at the
  csns back-n facility}, Chinese Physics B 31~(3) (2022) 038204.
\newblock \href {http://dx.doi.org/10.1088/1674-1056/ac48fd}
  {\path{doi:10.1088/1674-1056/ac48fd}}.
\newline\urlprefix\url{https://doi.org/10.1088/1674-1056/ac48fd}

\bibitem{FAN2020164343}
R.~Fan, H.~Jiang, W.~Jiang, G.~Zhang, H.~Yi, K.~Sun, H.~Bai, L.~Zhou, C.~Ning,
  Y.~He, Y.~Zhao, Y.~Wang, Y.~He, Z.~Cui, Z.~Sun, Q.~An, J.~Bao, P.~Cao,
  Q.~Chen, Y.~Chen, P.~Cheng, C.~Feng, M.~Gu, F.~Guo, C.~Han, Z.~Han, G.~He,
  H.~Huang, W.~Huang, X.~Huang, X.~Ji, X.~Ji, H.~Jing, L.~Kang, M.~Kang, B.~Li,
  L.~Li, Q.~Li, X.~Li, Y.~Li, Y.~Li, R.~Liu, S.~Liu, X.~Liu, G.~Luan, Y.~Ma,
  B.~Qi, J.~Ren, X.~Ruan, Z.~Song, H.~Sun, X.~Sun, Z.~Tan, H.~Tang, J.~Tang,
  P.~Wang, Q.~Wang, T.~Wang, Z.~Wang, Z.~Wang, J.~Wen, Z.~Wen, Q.~Wu, X.~Wu,
  X.~Wu, L.~Xie, Y.~Yang, L.~Yu, T.~Yu, Y.~Yu, J.~Zhang, L.~Zhang, L.~Zhang,
  Q.~Zhang, Q.~Zhang, X.~Zhang, Y.~Zhang, Z.~Zhang, Z.~Zhou, D.~Zhu, K.~Zhu,
  P.~Zhu,
  \href{https://www.sciencedirect.com/science/article/pii/S0168900220307403}{Detection
  of low-energy charged-particle using the de-e telescope at the back-n white
  neutron source}, Nuclear Instruments and Methods in Physics Research Section
  A: Accelerators, Spectrometers, Detectors and Associated Equipment 32 (2020)
  164343.
\newblock \href {http://dx.doi.org/https://doi.org/10.1016/j.nima.2020.164343}
  {\path{doi:https://doi.org/10.1016/j.nima.2020.164343}}.
\newline\urlprefix\url{https://www.sciencedirect.com/science/article/pii/S0168900220307403}

\bibitem{JIANG2020164126}
W.~Jiang, H.~Bai, H.~Jiang, H.~Yi, R.~Fan, G.~Zhang, J.~Tang, Z.~Sun, C.~Ning,
  K.~Sun, K.~Gao, Z.~Cui, Q.~An, J.~Bao, Y.~Bao, P.~Cao, H.~Chen, Q.~Chen,
  Y.~Chen, Y.~Chen, Z.~Chen, C.~Feng, M.~Gu, F.~Guo, C.~Han, Z.~Han, G.~He,
  Y.~He, Y.~Hong, H.~Huang, W.~Huang, X.~Huang, X.~Ji, X.~Ji, Z.~Jiang,
  H.~Jing, L.~Kang, M.~Kang, B.~Li, C.~Li, J.~Li, L.~Li, Q.~Li, X.~Li, Y.~Li,
  R.~Liu, S.~Liu, X.~Liu, G.~Luan, Q.~Mu, B.~Qi, J.~Ren, Z.~Ren, X.~Ruan,
  Y.~Song, Z.~Song, H.~Sun, X.~Sun, Z.~Tan, H.~Tang, X.~Tang, B.~Tian, L.~Wang,
  P.~Wang, Q.~Wang, T.~Wang, Y.~Wang, Z.~Wang, J.~Wen, Z.~Wen, Q.~Wu, X.~Wu,
  X.~Wu, L.~Xie, Y.~Yang, L.~Yu, T.~Yu, Y.~Yu, L.~Zhang, Q.~Zhang, X.~Zhang,
  Y.~Zhang, Z.~Zhang, Y.~Zhao, L.~Zhou, L.~Zhou, Z.~Zhou, D.~Zhu, K.~Zhu,
  P.~Zhu,
  \href{https://www.sciencedirect.com/science/article/pii/S016890022030526X}{Application
  of a silicon detector array in (n,lcp) reaction cross-section measurements at
  the csns back-n white neutron source}, Nuclear Instruments and Methods in
  Physics Research Section A: Accelerators, Spectrometers, Detectors and
  Associated Equipment 973 (2020) 164126.
\newblock \href {http://dx.doi.org/https://doi.org/10.1016/j.nima.2020.164126}
  {\path{doi:https://doi.org/10.1016/j.nima.2020.164126}}.
\newline\urlprefix\url{https://www.sciencedirect.com/science/article/pii/S016890022030526X}

\bibitem{Measurements2019}
H.~Jiang, W.~Jiang, H.~Bai, Z.~Cui, G.~Zhang, R.~Fan, H.~Yi, C.~Ning, L.~Zhou,
  J.~Tang, Q.~An, J.~Bao, Y.~Bao, P.~Cao, H.~Chen, Q.~Chen, Y.~Chen, Y.~Chen,
  Z.~Chen, C.~Feng, K.~Gao, M.~Gu, C.~Han, Z.~Han, G.~He, Y.~He, Y.~Hong,
  H.~Huang, W.~Huang, X.~Huang, X.~Ji, X.~Ji, Z.~Jiang, H.~Jing, L.~Kang,
  M.~Kang, B.~Li, C.~Li, J.~Li, L.~Li, Q.~Li, X.~Li, Y.~Li, R.~Liu, S.~Liu,
  X.~Liu, G.~Luan, Q.~Mu, B.~Qi, J.~Ren, Z.~Ren, X.~Ruan, Z.~Song, Y.~Song,
  H.~Sun, K.~Sun, X.~Sun, Z.~Sun, Z.~Tan, H.~Tang, X.~Tang, B.~Tian, L.~Wang,
  P.~Wang, Q.~Wang, T.~Wang, Z.~Wang, J.~Wen, Z.~Wen, Q.~Wu, X.~Wu, X.~Wu,
  L.~Xie, Y.~Yang, L.~Yu, T.~Yu, Y.~Yu, L.~Zhang, Q.~Zhang, X.~Zhang, Y.~Zhang,
  Z.~Zhang, Y.~Zhao, L.~Zhou, Z.~Zhou, D.~Zhu, K.~Zhu, P.~Zhu,
  \href{https://doi.org/10.1088/1674-1137/43/12/124002}{Measurements of
  differential and angle-integrated cross sections for the b10(n, a)li7
  reaction in the neutron energy range from 1.0 ev to 2.5 mev}, Chinese Physics
  C 43~(12) (2019) 124002.
\newblock \href {http://dx.doi.org/10.1088/1674-1137/43/12/124002}
  {\path{doi:10.1088/1674-1137/43/12/124002}}.
\newline\urlprefix\url{https://doi.org/10.1088/1674-1137/43/12/124002}

\bibitem{Bai_2020}
H.~Bai, R.~Fan, H.~Jiang, Z.~Cui, Y.~Hu, G.~Zhang, Z.~Chen, W.~Jiang, H.~Yi,
  J.~Tang, L.~Zhou, Q.~An, J.~Bao, P.~Cao, Q.~Chen, Y.~Chen, P.~Cheng, C.~Feng,
  M.~Gu, F.~Guo, C.~Han, Z.~Han, G.~He, Y.~He, Y.~He, H.~Huang, W.~Huang,
  X.~Huang, X.~Ji, X.~Ji, H.~Jing, L.~Kang, M.~Kang, B.~Li, L.~Li, Q.~Li,
  X.~Li, Y.~Li, Y.~Li, R.~Liu, S.~Liu, X.~Liu, G.~Luan, Y.~Ma, C.~Ning, B.~Qi,
  J.~Ren, X.~Ruan, Z.~Song, H.~Sun, X.~Sun, Z.~Sun, Z.~Tan, H.~Tang, P.~Wang,
  Q.~Wang, T.~Wang, Y.~Wang, Z.~Wang, Z.~Wang, J.~Wen, Z.~Wen, Q.~Wu, X.~Wu,
  X.~Wu, L.~Xie, Y.~Yang, L.~Yu, T.~Yu, Y.~Yu, J.~Zhang, L.~Zhang, L.~Zhang,
  Q.~Zhang, Q.~Zhang, X.~Zhang, Y.~Zhang, Z.~Zhang, Y.~Zhao, Z.~Zhou, D.~Zhu,
  K.~Zhu, P.~Zhu,
  \href{https://doi.org/10.1088/1674-1137/44/1/014003}{Measurement of the
  differential cross sections and angle-integrated cross sections of the li6(n,
  t)he4 reaction from 1.0 ev to 3.0 mev at the csns back-n white neutron
  source}, Chinese Physics C 44~(1) (2020) 014003.
\newblock \href {http://dx.doi.org/10.1088/1674-1137/44/1/014003}
  {\path{doi:10.1088/1674-1137/44/1/014003}}.
\newline\urlprefix\url{https://doi.org/10.1088/1674-1137/44/1/014003}

\bibitem{Cui2021}
Z.~{Cui}, H.~{Jiang}, W.~{Jiang}, G.~{Zhang}, R.~{Fan}, D.~{Pang}, K.~{Sun},
  H.~{Bai}, Y.~{Hu}, J.~{Liu}, H.~{Yi}, C.~{Ning}, Z.~{Sun}, J.~{Tang},
  Q.~{An}, J.~{Bao}, Y.~{Bao}, P.~{Cao}, H.~{Chen}, Q.~{Chen}, Y.~{Chen},
  Y.~{Chen}, Z.~{Chen}, C.~{Feng}, K.~{Gao}, M.~{Gu}, C.~{Han}, Z.~{Han},
  G.~{He}, Y.~{He}, Y.~{Hong}, H.~{Huang}, W.~{Huang}, X.~{Huang}, X.~{Ji},
  X.~{Ji}, Z.~{Jiang}, H.~{Jing}, L.~{Kang}, M.~{Kang}, B.~{Li}, C.~{Li},
  J.~{Li}, L.~{Li}, Q.~{Li}, X.~{Li}, Y.~{Li}, R.~{Liu}, S.~{Liu}, X.~{Liu},
  G.~{Luan}, Q.~{Mu}, B.~{Qi}, J.~{Ren}, Z.~{Ren}, X.~{Ruan}, Z.~{Song},
  Y.~{Song}, H.~{Sun}, X.~{Sun}, Z.~{Tan}, H.~{Tang}, X.~{Tang}, B.~{Tian},
  L.~{Wang}, P.~{Wang}, Q.~{Wang}, T.~{Wang}, Z.~{Wang}, J.~{Wen}, Z.~{Wen},
  Q.~{Wu}, X.~{Wu}, X.~{Wu}, L.~{Xie}, Y.~{Yang}, L.~{Yu}, T.~{Yu}, Y.~{Yu},
  L.~{Zhang}, Q.~{Zhang}, X.~{Zhang}, Y.~{Zhang}, Z.~{Zhang}, Y.~{Zhao},
  L.~{Zhou}, Z.~{Zhou}, D.~{Zhu}, K.~{Zhu}, P.~{Zhu}, {CSNS Back-n
  Collaboration}, {Measurement of relative differential cross sections of the
  neutron-deuteron elastic scattering for neutron energy from 13 to 52 MeV},
  European Physical Journal A 57~(11) (2021) 310.
\newblock \href {http://arxiv.org/abs/2112.03520} {\path{arXiv:2112.03520}},
  \href {http://dx.doi.org/10.1140/epja/s10050-021-00610-9}
  {\path{doi:10.1140/epja/s10050-021-00610-9}}.

\bibitem{Measurement2021}
Z.~Cui, H.~Jiang, K.~Sun, G.~Zhang, R.~Fan, W.~Jiang, H.~Bai, Y.~Hu, J.~Liu,
  H.~Yi, C.~Ning, Z.~Sun, J.~Tang, Q.~An, J.~Bao, Y.~Bao, P.~Cao, H.~Chen,
  Q.~Chen, Y.~Chen, Y.~Chen, Z.~Chen, C.~Feng, K.~Gao, M.~Gu, C.~Han, Z.~Han,
  G.~He, Y.~He, Y.~Hong, H.~Huang, W.~Huang, X.~Huang, X.~Ji, X.~Ji, Z.~Jiang,
  H.~Jing, L.~Kang, M.~Kang, B.~Li, C.~Li, J.~Li, L.~Li, Q.~Li, X.~Li, Y.~Li,
  R.~Liu, S.~Liu, X.~Liu, G.~Luan, Q.~Mu, B.~Qi, J.~Ren, Z.~Ren, X.~Ruan,
  Z.~Song, Y.~Song, H.~Sun, X.~Sun, Z.~Tan, H.~Tang, X.~Tang, B.~Tian, L.~Wang,
  P.~Wang, Q.~Wang, T.~Wang, Z.~Wang, J.~Wen, Z.~Wen, Q.~Wu, X.~Wu, X.~Wu,
  L.~Xie, Y.~Yang, L.~Yu, T.~Yu, Y.~Yu, L.~Zhang, Q.~Zhang, X.~Zhang, Y.~Zhang,
  Z.~Zhang, Y.~Zhao, L.~Zhou, Z.~Zhou, D.~Zhu, K.~Zhu, P.~Zhu, ({The CSNS
  Back-n Collaboration}),
  \href{https://doi.org/10.1088/1674-1137/45/6/064001}{Measurement of
  differential cross sections of the neutron-induced deuteron production
  reactions on carbon from 25 to 52 mev}, Chinese Physics C 45 (2021) 064001.
\newblock \href {http://dx.doi.org/10.1088/1674-1137/abf136}
  {\path{doi:10.1088/1674-1137/abf136}}.
\newline\urlprefix\url{https://doi.org/10.1088/1674-1137/45/6/064001}

\bibitem{Chen2019}
Y.~Chen, G.~Luan, J.~Bao, H.~Jing, L.~Zhang, Q.~An, H.~Bai, P.~Cao, Q.~Chen,
  P.~Cheng, Z.~Cui, R.~Fan, C.~Feng, M.~Gu, F.~Guo, C.~Han, Z.~Han, G.~He,
  Y.~He, Y.~He, H.~Huang, W.~Huang, X.~Huang, X.~Ji, X.~Ji, H.~Jiang, W.~Jiang,
  L.~Kang, M.~Kang, B.~Li, L.~Li, Q.~Li, X.~Li, Y.~Li, Y.~Li, R.~Liu, S.~Liu,
  X.~Liu, Y.~Ma, C.~Ning, B.~Qi, J.~Ren, X.~Ruan, Z.~Song, H.~Sun, X.~Sun,
  Z.~Sun, Z.~Tan, H.~Tang, J.~Tang, P.~Wang, Q.~Wang, T.~Wang, Y.~Wang,
  Z.~Wang, Z.~Wang, J.~Wen, Z.~Wen, Q.~Wu, X.~Wu, X.~Wu, L.~Xie, Y.~Yang,
  H.~Yi, L.~Yu, T.~Yu, Y.~Yu, G.~Zhang, J.~Zhang, L.~Zhang, Q.~Zhang, Q.~Zhang,
  X.~Zhang, Y.~Zhang, Z.~Zhang, Y.~Zhao, L.~Zhou, Z.~Zhou, D.~Zhu, K.~Zhu,
  P.~Zhu, Neutron energy spectrum measurement of the back-n white neutron
  source at csns, The European Physical Journal A 55 (2019) 1--10.

\bibitem{An_2017}
Q.~An, H.~Bai, J.~Bao, P.~Cao, Y.~Chen, Y.~Chen, P.~Cheng, R.~Fan, C.~Feng,
  J.~Gu, M.~Gu, B.~He, G.~He, W.~He, Y.~He, Y.~He, H.~Huang, X.~Huang,
  W.~Huang, X.~Ji, X.~Ji, H.~Jing, B.~Li, C.~Li, G.~Li, Q.~Li, Y.~Li, R.~Liu,
  S.~Liu, G.~Luan, Y.~L. Ma, M.~Peng, C.~Ning, X.~Qi, J.~Ren, X.~Ruan, B.~Shi,
  Z.~Song, X.~Su, Z.~Sun, H.~Tang, J.~Tang, Z.~Tan, P.~Wang, Q.~Wang, Q.~Wang,
  Y.~F. Wang, Z.~Wang, J.~Wen, Z.~Wen, Q.~Wu, X.~Wu, Y.~Yang, T.~Yu, Y.~Yu,
  D.~Zhang, G.~Zhang, H.~Zhang, J.~Zhang, L.~Zhang, Q.~Zhang, X.~Zhang,
  Y.~Zhang, Y.~Zhao, L.~Zheng, Y.~Zheng, J.~Zhong, Q.~Zhong, L.~Zhou, Z.~Zhou,
  K.~Zhu, \href{https://doi.org/10.1088/1748-0221/12/07/p07022}{Back-n white
  neutron facility for nuclear data measurements at {CSNS}}, Journal of
  Instrumentation 12~(07) (2017) P07022--P07022.
\newblock \href {http://dx.doi.org/10.1088/1748-0221/12/07/p07022}
  {\path{doi:10.1088/1748-0221/12/07/p07022}}.
\newline\urlprefix\url{https://doi.org/10.1088/1748-0221/12/07/p07022}

\bibitem{Geant4}
S.~Agostinelli, J.~Allison, K.~Amako, J.~Apostolakis, H.~Araujo, P.~Arce,
  M.~Asai, D.~Axen, S.~Banerjee, G.~Barrand, F.~Behner, L.~Bellagamba,
  J.~Boudreau, L.~Broglia, A.~Brunengo, H.~Burkhardt, S.~Chauvie, J.~Chuma,
  R.~Chytracek, G.~Cooperman, G.~Cosmo, P.~Degtyarenko, A.~Dell'Acqua,
  G.~Depaola, D.~Dietrich, R.~Enami, A.~Feliciello, C.~Ferguson, H.~Fesefeldt,
  G.~Folger, F.~Foppiano, A.~Forti, S.~Garelli, S.~Giani, R.~Giannitrapani,
  D.~Gibin, J.~{Gómez Cadenas}, I.~González, G.~{Gracia Abril}, G.~Greeniaus,
  W.~Greiner, V.~Grichine, A.~Grossheim, S.~Guatelli, P.~Gumplinger,
  R.~Hamatsu, K.~Hashimoto, H.~Hasui, A.~Heikkinen, A.~Howard, V.~Ivanchenko,
  A.~Johnson, F.~Jones, J.~Kallenbach, N.~Kanaya, M.~Kawabata, Y.~Kawabata,
  M.~Kawaguti, S.~Kelner, P.~Kent, A.~Kimura, T.~Kodama, R.~Kokoulin,
  M.~Kossov, H.~Kurashige, E.~Lamanna, T.~Lampén, V.~Lara, V.~Lefebure,
  F.~Lei, M.~Liendl, W.~Lockman, F.~Longo, S.~Magni, M.~Maire, E.~Medernach,
  K.~Minamimoto, P.~{Mora de Freitas}, Y.~Morita, K.~Murakami, M.~Nagamatu,
  R.~Nartallo, P.~Nieminen, T.~Nishimura, K.~Ohtsubo, M.~Okamura, S.~O'Neale,
  Y.~Oohata, K.~Paech, J.~Perl, A.~Pfeiffer, M.~Pia, F.~Ranjard, A.~Rybin,
  S.~Sadilov, E.~{Di Salvo}, G.~Santin, T.~Sasaki, N.~Savvas, Y.~Sawada,
  S.~Scherer, S.~Sei, V.~Sirotenko, D.~Smith, N.~Starkov, H.~Stoecker,
  J.~Sulkimo, M.~Takahata, S.~Tanaka, E.~Tcherniaev, E.~{Safai Tehrani},
  M.~Tropeano, P.~Truscott, H.~Uno, L.~Urban, P.~Urban, M.~Verderi, A.~Walkden,
  W.~Wander, H.~Weber, J.~Wellisch, T.~Wenaus, D.~Williams, D.~Wright,
  T.~Yamada, H.~Yoshida, D.~Zschiesche,
  \href{https://www.sciencedirect.com/science/article/pii/S0168900203013688}{Geant4—a
  simulation toolkit}, Nuclear Instruments and Methods in Physics Research
  Section A: Accelerators, Spectrometers, Detectors and Associated Equipment
  506~(3) (2003) 250--303.
\newblock \href
  {http://dx.doi.org/https://doi.org/10.1016/S0168-9002(03)01368-8}
  {\path{doi:https://doi.org/10.1016/S0168-9002(03)01368-8}}.
\newline\urlprefix\url{https://www.sciencedirect.com/science/article/pii/S0168900203013688}

\bibitem{Yi_2020}
H.~Yi, T.~Wang, Y.~Li, X.~Ruan, J.~Ren, Y.~Chen, Q.~Li, J.~Wen, J.~Tang, Q.~An,
  H.~Bai, J.~Bao, Y.~Bao, P.~Cao, H.~Chen, Q.~Chen, Y.~Chen, Z.~Chen, Z.~Cui,
  R.~Fan, C.~Feng, K.~Gao, M.~Gu, C.~Han, Z.~Han, G.~He, Y.~He, Y.~Hong,
  H.~Huang, W.~Huang, X.~Huang, X.~Ji, X.~Ji, H.~Jiang, W.~Jiang, Z.~Jiang,
  H.~Jing, L.~Kang, M.~Kang, B.~Li, C.~Li, J.~Li, L.~Li, X.~Li, R.~Liu, S.~Liu,
  X.~Liu, G.~Luan, Q.~Mu, C.~Ning, B.~Qi, Z.~Ren, Y.~Song, Z.~Song, H.~Sun,
  K.~Sun, X.~Sun, Z.~Sun, Z.~Tan, H.~Tang, X.~Tang, B.~Tian, L.~Wang, P.~Wang,
  Q.~Wang, Z.~Wang, Z.~Wen, Q.~Wu, X.~Wu, X.~Wu, L.~Xie, Y.~Yang, L.~Yu, T.~Yu,
  Y.~Yu, G.~Zhang, L.~Zhang, Q.~Zhang, X.~Zhang, Y.~Zhang, Z.~Zhang, Y.~Zhao,
  L.~Zhou, Z.~Zhou, D.~Zhu, K.~Zhu, P.~Zhu,
  \href{https://doi.org/10.1088/1748-0221/15/03/p03026}{Double-bunch unfolding
  methods for the back-n white neutron source at csns}, Journal of
  Instrumentation 15~(03) (2020) P03026--P03026.
\newblock \href {http://dx.doi.org/10.1088/1748-0221/15/03/p03026}
  {\path{doi:10.1088/1748-0221/15/03/p03026}}.
\newline\urlprefix\url{https://doi.org/10.1088/1748-0221/15/03/p03026}

\bibitem{KONING20122841}
A.~Koning, D.~Rochman,
  \href{https://www.sciencedirect.com/science/article/pii/S0090375212000889}{Modern
  nuclear data evaluation with the talys code system}, Nuclear Data Sheets
  113~(12) (2012) 2841--2934, special Issue on Nuclear Reaction Data.
\newblock \href {http://dx.doi.org/https://doi.org/10.1016/j.nds.2012.11.002}
  {\path{doi:https://doi.org/10.1016/j.nds.2012.11.002}}.
\newline\urlprefix\url{https://www.sciencedirect.com/science/article/pii/S0090375212000889}

\bibitem{BROWN20181}
D.~Brown, M.~Chadwick, R.~Capote, A.~Kahler, A.~Trkov, M.~Herman, A.~Sonzogni,
  Y.~Danon, A.~Carlson, M.~Dunn, D.~Smith, G.~Hale, G.~Arbanas, R.~Arcilla,
  C.~Bates, B.~Beck, B.~Becker, F.~Brown, R.~Casperson, J.~Conlin, D.~Cullen,
  M.-A. Descalle, R.~Firestone, T.~Gaines, K.~Guber, A.~Hawari, J.~Holmes,
  T.~Johnson, T.~Kawano, B.~Kiedrowski, A.~Koning, S.~Kopecky, L.~Leal,
  J.~Lestone, C.~Lubitz, J.~{Márquez Damián}, C.~Mattoon, E.~McCutchan,
  S.~Mughabghab, P.~Navratil, D.~Neudecker, G.~Nobre, G.~Noguere, M.~Paris,
  M.~Pigni, A.~Plompen, B.~Pritychenko, V.~Pronyaev, D.~Roubtsov, D.~Rochman,
  P.~Romano, P.~Schillebeeckx, S.~Simakov, M.~Sin, I.~Sirakov, B.~Sleaford,
  V.~Sobes, E.~Soukhovitskii, I.~Stetcu, P.~Talou, I.~Thompson, S.~{van der
  Marck}, L.~Welser-Sherrill, D.~Wiarda, M.~White, J.~Wormald, R.~Wright,
  M.~Zerkle, G.~Žerovnik, Y.~Zhu,
  \href{https://www.sciencedirect.com/science/article/pii/S0090375218300206}{Endf/b-viii.0:
  The 8th major release of the nuclear reaction data library with cielo-project
  cross sections, new standards and thermal scattering data}, Nuclear Data
  Sheets 148 (2018) 1--142, special Issue on Nuclear Reaction Data.
\newblock \href {http://dx.doi.org/https://doi.org/10.1016/j.nds.2018.02.001}
  {\path{doi:https://doi.org/10.1016/j.nds.2018.02.001}}.
\newline\urlprefix\url{https://www.sciencedirect.com/science/article/pii/S0090375218300206}

\bibitem{PhysRevC.104.014603}
S.~A. Kuvin, H.~Y. Lee, B.~DiGiovine, A.~Georgiadou, S.~Mosby, D.~Votaw,
  M.~White, L.~Zavorka,
  \href{https://link.aps.org/doi/10.1103/PhysRevC.104.014603}{Validation of
  neutron-induced reactions on natural carbon using an active target at neutron
  energies up to 22 mev at lansce}, Phys. Rev. C 104 (2021) 014603.
\newblock \href {http://dx.doi.org/10.1103/PhysRevC.104.014603}
  {\path{doi:10.1103/PhysRevC.104.014603}}.
\newline\urlprefix\url{https://link.aps.org/doi/10.1103/PhysRevC.104.014603}

\bibitem{KONING2003231}
A.~Koning, J.~Delaroche,
  \href{https://www.sciencedirect.com/science/article/pii/S0375947402013210}{Local
  and global nucleon optical models from 1 kev to 200 mev}, Nuclear Physics A
  713~(3) (2003) 231--310.
\newblock \href
  {http://dx.doi.org/https://doi.org/10.1016/S0375-9474(02)01321-0}
  {\path{doi:https://doi.org/10.1016/S0375-9474(02)01321-0}}.
\newline\urlprefix\url{https://www.sciencedirect.com/science/article/pii/S0375947402013210}

\bibitem{PEREY19761}
C.~Perey, F.~Perey,
  \href{https://www.sciencedirect.com/science/article/pii/0092640X76900073}{Compilation
  of phenomenological optical-model parameters 1954–1975}, Atomic Data and
  Nuclear Data Tables 17~(1) (1976) 1--101.
\newblock \href
  {http://dx.doi.org/https://doi.org/10.1016/0092-640X(76)90007-3}
  {\path{doi:https://doi.org/10.1016/0092-640X(76)90007-3}}.
\newline\urlprefix\url{https://www.sciencedirect.com/science/article/pii/0092640X76900073}

\bibitem{KARIEM2006}
S.~Abdel-Kariem,
  \href{https://journals.tubitak.gov.tr/physics/vol30/iss1/1}{Dwba of the
  reaction $^{9}be(p,\alpha)^{6}li$ at $e_p = 18.6\sim 50 mev$}, Turkish
  Journal of Physics 30~(1) (2006) 1--14.
\newline\urlprefix\url{https://journals.tubitak.gov.tr/physics/vol30/iss1/1}

\bibitem{KURATH1975269}
D.~Kurath, D.~Millener,
  \href{https://www.sciencedirect.com/science/article/pii/037594747590353X}{Three-nucleon
  transfer for the 1p shell}, Nuclear Physics A 238~(2) (1975) 269--286.
\newblock \href
  {http://dx.doi.org/https://doi.org/10.1016/0375-9474(75)90353-X}
  {\path{doi:https://doi.org/10.1016/0375-9474(75)90353-X}}.
\newline\urlprefix\url{https://www.sciencedirect.com/science/article/pii/037594747590353X}

\bibitem{JENNY2007}
J.~Lee, M.~B. Tsang, W.~G. Lynch,
  \href{https://link.aps.org/doi/10.1103/PhysRevC.75.064320}{Neutron
  spectroscopic factors from transfer reactions}, Phys. Rev. C 75 (2007)
  064320.
\newblock \href {http://dx.doi.org/10.1103/PhysRevC.75.064320}
  {\path{doi:10.1103/PhysRevC.75.064320}}.
\newline\urlprefix\url{https://link.aps.org/doi/10.1103/PhysRevC.75.064320}

\bibitem{SINHA1930}
B.~B.~P. Sinha, G.~A. Peterson, R.~R. Whitney, I.~Sick, J.~S. McCarthy,
  \href{https://link.aps.org/doi/10.1103/PhysRevC.7.1930}{Nuclear charge
  distributions of isotone pairs. ii. $^{39}\mathrm{K}$ and
  $^{40}\mathrm{Ca}$}, Phys. Rev. C 7 (1973) 1930--1938.
\newblock \href {http://dx.doi.org/10.1103/PhysRevC.7.1930}
  {\path{doi:10.1103/PhysRevC.7.1930}}.
\newline\urlprefix\url{https://link.aps.org/doi/10.1103/PhysRevC.7.1930}

\bibitem{DEVRIES1972424}
R.~Devries, J.-L. Perrenoud, I.~Slaus, J.~Sunier,
  \href{https://www.sciencedirect.com/science/article/pii/0375947472904708}{Finite-range
  dwba analysis of $(p, \alpha)$ reactions on $^{9}be$ and $^{11}b$}, Nuclear
  Physics A 178~(2) (1972) 424--436.
\newblock \href
  {http://dx.doi.org/https://doi.org/10.1016/0375-9474(72)90470-8}
  {\path{doi:https://doi.org/10.1016/0375-9474(72)90470-8}}.
\newline\urlprefix\url{https://www.sciencedirect.com/science/article/pii/0375947472904708}

\bibitem{ESWARAN1990}
M.~A. Eswaran, S.~Kumar, E.~T. Mirgule,
  \href{https://link.aps.org/doi/10.1103/PhysRevC.42.1036}{Direct cluster
  transfer in $^{3}\mathrm{He}$ bombardment of $^{13}\mathrm{C}$ at sub-barrier
  energies}, Phys. Rev. C 42 (1990) 1036--1042.
\newblock \href {http://dx.doi.org/10.1103/PhysRevC.42.1036}
  {\path{doi:10.1103/PhysRevC.42.1036}}.
\newline\urlprefix\url{https://link.aps.org/doi/10.1103/PhysRevC.42.1036}

\end{thebibliography}

\appendix
\counterwithin{figure}{section}
\counterwithin{table}{section}
\begin{appendices}

\section{Appendix A}

\begin{longtable}[width=0.9\linewidth,cols=8,pos=htb]{@{} LLLLLLLL@{}}
	\caption{Differential cross sections of the $^{12}C(n,\alpha)x$ reactions measured at $\theta_{L} = 24.5^{\circ}, 34.0^{\circ}, 43.5^{\circ}, 53.0^{\circ}, 62.5^{\circ}, 72.0^{\circ} and 81.5 ^{\circ}$.(Only statistical uncertainties were included) }\label{macs_t}\\
	\toprule 
	$E_{n}$(MeV)& $24.5^{\circ}$(mb) & $34.0^{\circ}$(mb) & $43.5^{\circ}$(mb) & $53.0^{\circ}$(mb) & $62.5^{\circ}$(mb)& $72.0^{\circ}$(mb) & $81.5^{\circ}$(mb)\\ 
	\midrule 
	\endfirsthead
	
	\multicolumn{8}{l}{{\bfseries \tablename\ \thetable{}--continued from previous page}}\\
	\toprule 
	$E_{n}$(MeV)& $24.5^{\circ}$(mb) & $34.0^{\circ}$(mb) & $43.5^{\circ}$(mb) & $53.0^{\circ}$(mb) & $62.5^{\circ}$(mb)& $72.0^{\circ}$(mb) & $81.5^{\circ}$(mb)\\ 
	\midrule
	\endhead 
	
	\bottomrule 
	\multicolumn{8}{l}{{Continued on next page}}\\
	\endfoot 
	
	\hline 
	\endlastfoot
        6.2&      0.00$\pm$0.02&      0.00$\pm$0.01&      0.01$\pm$0.03&      0.00$\pm$0.02&      0.00$\pm$0.02&      0.00$\pm$0.01&      0.00$\pm$0.00\\
       6.5&      1.79$\pm$0.35&      0.63$\pm$0.20&      0.77$\pm$0.22&      0.13$\pm$0.09&      0.31$\pm$0.15&      0.03$\pm$0.05&      0.22$\pm$0.12\\
       6.8&      2.71$\pm$0.43&      1.23$\pm$0.28&      0.19$\pm$0.11&      0.61$\pm$0.20&      0.35$\pm$0.16&      0.09$\pm$0.08&      0.01$\pm$0.03\\
       7.2&      0.25$\pm$0.13&      0.11$\pm$0.08&      0.15$\pm$0.09&      0.41$\pm$0.16&      0.23$\pm$0.13&      0.02$\pm$0.04&      0.03$\pm$0.04\\
       7.6&      0.94$\pm$0.26&      0.19$\pm$0.11&      0.19$\pm$0.11&      0.00$\pm$0.02&      0.31$\pm$0.15&      0.10$\pm$0.08&      0.00$\pm$0.01\\
       8.0&      1.88$\pm$0.36&      0.61$\pm$0.20&      0.91$\pm$0.24&      0.00$\pm$0.00&      0.11$\pm$0.09&      0.01$\pm$0.02&      0.01$\pm$0.03\\
       8.5&      3.38$\pm$0.49&      1.81$\pm$0.35&      2.28$\pm$0.38&      0.48$\pm$0.18&      0.68$\pm$0.23&      0.12$\pm$0.09&      0.20$\pm$0.12\\
       9.0&     13.05$\pm$0.98&      5.76$\pm$0.63&      9.14$\pm$0.78&      1.62$\pm$0.34&      4.21$\pm$0.57&      0.78$\pm$0.24&      1.20$\pm$0.30\\
       9.5&     40.74$\pm$1.78&     19.62$\pm$1.19&      7.30$\pm$0.72&      8.90$\pm$0.82&      5.08$\pm$0.64&      3.95$\pm$0.56&      2.58$\pm$0.45\\
      10.1&     15.16$\pm$1.10&     12.67$\pm$0.97&      2.33$\pm$0.41&      7.80$\pm$0.78&      1.52$\pm$0.36&      1.85$\pm$0.39&      0.24$\pm$0.14\\
      10.8&      5.53$\pm$0.68&      6.88$\pm$0.73&      1.61$\pm$0.35&      2.69$\pm$0.46&      0.48$\pm$0.21&      0.36$\pm$0.17&      0.04$\pm$0.06\\
      11.5&     11.17$\pm$0.98&     10.76$\pm$0.93&      4.20$\pm$0.57&      5.36$\pm$0.67&      1.60$\pm$0.38&      0.51$\pm$0.21&      0.34$\pm$0.17\\
      12.3&     33.73$\pm$1.73&     18.46$\pm$1.24&     20.70$\pm$1.29&      7.19$\pm$0.78&      7.02$\pm$0.81&      3.14$\pm$0.53&      3.13$\pm$0.53\\
      13.2&     91.23$\pm$2.93&     37.82$\pm$1.82&     53.32$\pm$2.13&     10.33$\pm$0.97&     27.67$\pm$1.65&     10.57$\pm$1.00&     12.61$\pm$1.10\\
      14.2&     94.25$\pm$3.02&     48.57$\pm$2.09&     51.22$\pm$2.12&      9.46$\pm$0.94&     30.95$\pm$1.77&     27.24$\pm$1.63&     17.40$\pm$1.30\\
      15.4&     80.02$\pm$2.83&     36.26$\pm$1.84&     44.00$\pm$1.99&     13.72$\pm$1.15&     27.90$\pm$1.71&     27.15$\pm$1.65&     16.31$\pm$1.28\\
      16.6&     45.58$\pm$2.16&     32.52$\pm$1.76&     28.86$\pm$1.63&     12.68$\pm$1.12&     17.81$\pm$1.38&     16.88$\pm$1.31&     11.33$\pm$1.08\\
      18.0&     55.41$\pm$2.40&     46.71$\pm$2.13&     28.73$\pm$1.64&     27.40$\pm$1.66&     16.65$\pm$1.34&     10.15$\pm$1.03&     10.67$\pm$1.06\\
      19.7&     57.30$\pm$2.46&     59.86$\pm$2.43&     37.25$\pm$1.89&     50.03$\pm$2.25&     20.64$\pm$1.51&      9.42$\pm$1.00&      9.47$\pm$1.01\\
      21.5&     87.56$\pm$3.07&     73.45$\pm$2.71&     45.96$\pm$2.11&     50.74$\pm$2.29&     23.51$\pm$1.62&     11.72$\pm$1.12&     13.12$\pm$1.19\\
      23.6&    100.44$\pm$3.29&     79.76$\pm$2.83&     48.03$\pm$2.16&     47.19$\pm$2.21&     28.68$\pm$1.80&     18.76$\pm$1.42&     13.52$\pm$1.21\\
      26.1&     94.57$\pm$3.22&     69.52$\pm$2.66&     61.01$\pm$2.46&     45.72$\pm$2.20&     29.63$\pm$1.84&     16.48$\pm$1.35&     15.67$\pm$1.32\\
      29.0&    114.57$\pm$3.58&     77.04$\pm$2.83&     74.35$\pm$2.74&     48.41$\pm$2.28&     43.14$\pm$2.24&     25.65$\pm$1.70&     22.11$\pm$1.58\\
      32.4&    110.55$\pm$3.53&     82.03$\pm$2.94&     64.08$\pm$2.56&     46.39$\pm$2.24&     39.08$\pm$2.15&     37.87$\pm$2.07&     24.53$\pm$1.67\\
      36.4&     63.80$\pm$2.63&     52.21$\pm$2.30&     36.33$\pm$1.89&     35.88$\pm$1.94&     24.92$\pm$1.68&     18.30$\pm$1.41&      9.89$\pm$1.04\\
      41.3&     40.95$\pm$2.12&     38.06$\pm$1.97&     15.69$\pm$1.25&     25.16$\pm$1.63&     14.14$\pm$1.27&      7.10$\pm$0.88&      7.19$\pm$0.89\\
      47.2&     27.38$\pm$1.73&     31.79$\pm$1.80&     18.00$\pm$1.34&     21.01$\pm$1.49&     12.94$\pm$1.22&      2.84$\pm$0.56&      6.36$\pm$0.84\\
      54.6&     40.93$\pm$2.13&     32.66$\pm$1.83&     35.10$\pm$1.87&     17.52$\pm$1.36&     20.22$\pm$1.53&     10.61$\pm$1.08&      7.49$\pm$0.91\\
      63.9&     64.16$\pm$2.72&     45.80$\pm$2.22&     48.72$\pm$2.25&     29.39$\pm$1.80&     30.14$\pm$1.90&     16.79$\pm$1.39&     13.77$\pm$1.27\\
      76.0&     60.93$\pm$2.75&     44.10$\pm$2.26&     39.89$\pm$2.12&     28.92$\pm$1.86&     26.96$\pm$1.87&     19.49$\pm$1.56&     13.40$\pm$1.30\\
	\bottomrule	
	
\end{longtable}

\begin{longtable}[width=0.9\linewidth,cols=7,pos=htb]{@{} LLLLLLL@{}}
	\caption{Differential cross sections of the $^{12}C(n,\alpha)x$ reactions measured at $\theta_{L} = 91.0^{\circ},\theta_{R} = 98.5^{\circ}, 108.0^{\circ}, 127.0^{\circ}, 136.5^{\circ} and 146.0 ^{\circ}$. (Only statistical uncertainties were included) }\label{macs2_t}\\
	\toprule 
	$E_{n}$(MeV)& $91.0^{\circ}$(mb) & $98.5^{\circ}$(mb) & $108.0^{\circ}$(mb) & $127.0^{\circ}$(mb) & $136.5^{\circ}$(mb)& $146.0^{\circ}$(mb)\\ 
	\midrule 
	\endfirsthead
	
	\multicolumn{7}{l}{{\bfseries \tablename\ \thetable{}--continued from previous page}}\\
	\toprule 
	$E_{n}$(MeV)& $91.0^{\circ}$(mb) & $98.5^{\circ}$(mb) & $108.0^{\circ}$(mb) & $127.0^{\circ}$(mb) & $136.5^{\circ}$(mb)& $146.0^{\circ}$(mb)\\  \midrule
	\endhead 
	\bottomrule 
	\multicolumn{7}{l}{{Continued on next page}}\\
	\endfoot 
	
	\hline 
	\endlastfoot
           6.2&      0.01$\pm$0.02&      0.00$\pm$0.00&      0.00$\pm$0.01&      0.00$\pm$0.00&      0.00$\pm$0.00&      0.00$\pm$0.00\\
       6.5&      0.06$\pm$0.07&      0.34$\pm$0.15&      0.02$\pm$0.04&      0.04$\pm$0.05&      0.00$\pm$0.02&      0.04$\pm$0.05\\
       6.8&      0.13$\pm$0.10&      0.01$\pm$0.02&      0.01$\pm$0.03&      0.04$\pm$0.05&      0.00$\pm$0.01&      0.01$\pm$0.02\\
       7.2&      0.01$\pm$0.02&      0.00$\pm$0.00&      0.00$\pm$0.01&      0.02$\pm$0.03&      0.00$\pm$0.00&      0.00$\pm$0.01\\
       7.6&      0.12$\pm$0.10&      0.10$\pm$0.08&      0.00$\pm$0.00&      0.09$\pm$0.08&      0.00$\pm$0.00&      0.00$\pm$0.00\\
       8.0&      0.08$\pm$0.08&      0.00$\pm$0.00&      0.00$\pm$0.00&      0.00$\pm$0.00&      0.00$\pm$0.00&      0.00$\pm$0.00\\
       8.5&      0.04$\pm$0.06&      0.15$\pm$0.10&      0.15$\pm$0.10&      0.01$\pm$0.02&      0.27$\pm$0.14&      0.03$\pm$0.04\\
       9.0&      1.23$\pm$0.32&      2.99$\pm$0.45&      3.25$\pm$0.47&      0.80$\pm$0.24&      1.15$\pm$0.29&      1.56$\pm$0.32\\
       9.5&      3.40$\pm$0.55&      3.30$\pm$0.49&      2.26$\pm$0.40&      1.33$\pm$0.31&      2.32$\pm$0.42&      1.40$\pm$0.32\\
      10.1&      2.14$\pm$0.44&      0.23$\pm$0.13&      0.51$\pm$0.20&      0.26$\pm$0.14&      0.07$\pm$0.08&      0.29$\pm$0.15\\
      10.8&      0.94$\pm$0.30&      0.20$\pm$0.12&      0.09$\pm$0.08&      0.03$\pm$0.05&      0.00$\pm$0.01&      0.02$\pm$0.04\\
      11.5&      0.42$\pm$0.20&      0.45$\pm$0.19&      0.55$\pm$0.21&      0.24$\pm$0.14&      0.36$\pm$0.17&      0.34$\pm$0.16\\
      12.3&      0.81$\pm$0.29&      0.64$\pm$0.23&      1.39$\pm$0.34&      0.65$\pm$0.23&      0.15$\pm$0.11&      0.53$\pm$0.21\\
      13.2&      4.20$\pm$0.67&      0.94$\pm$0.29&      1.93$\pm$0.41&      1.30$\pm$0.34&      1.21$\pm$0.34&      0.78$\pm$0.26\\
      14.2&      5.90$\pm$0.81&      3.37$\pm$0.55&      4.04$\pm$0.60&      2.71$\pm$0.50&      1.83$\pm$0.42&      1.66$\pm$0.38\\
      15.4&      4.85$\pm$0.74&      1.11$\pm$0.32&      2.99$\pm$0.53&      2.86$\pm$0.52&      2.61$\pm$0.51&      1.05$\pm$0.31\\
      16.6&      5.30$\pm$0.78&      2.43$\pm$0.48&      3.78$\pm$0.60&      1.92$\pm$0.43&      1.86$\pm$0.43&      2.16$\pm$0.45\\
      18.0&      7.04$\pm$0.91&      5.52$\pm$0.73&      3.89$\pm$0.61&      2.74$\pm$0.52&      4.63$\pm$0.69&      4.86$\pm$0.68\\
      19.7&     10.63$\pm$1.13&      8.02$\pm$0.88&      7.11$\pm$0.83&      4.59$\pm$0.68&      3.59$\pm$0.61&      5.14$\pm$0.70\\
      21.5&      7.34$\pm$0.95&     13.22$\pm$1.15&      8.48$\pm$0.92&      6.63$\pm$0.82&      7.17$\pm$0.87&      7.65$\pm$0.86\\
      23.6&     14.58$\pm$1.34&     15.46$\pm$1.24&     10.17$\pm$1.01&      8.64$\pm$0.94&     11.09$\pm$1.09&      9.00$\pm$0.94\\
      26.1&     13.52$\pm$1.30&     13.93$\pm$1.19&     11.24$\pm$1.07&      9.42$\pm$0.99&     10.95$\pm$1.09&      9.38$\pm$0.97\\
      29.0&     15.09$\pm$1.39&     11.48$\pm$1.09&     10.36$\pm$1.04&     10.31$\pm$1.05&      8.30$\pm$0.96&      7.89$\pm$0.90\\
      32.4&     13.16$\pm$1.30&      9.02$\pm$0.97&      9.77$\pm$1.01&      7.26$\pm$0.88&      6.88$\pm$0.88&      4.38$\pm$0.67\\
      36.4&     11.94$\pm$1.22&      4.18$\pm$0.65&      4.16$\pm$0.65&      3.84$\pm$0.63&      4.18$\pm$0.67&      1.95$\pm$0.44\\
      41.3&      6.67$\pm$0.91&      5.06$\pm$0.72&      3.88$\pm$0.63&      2.36$\pm$0.50&      1.60$\pm$0.42&      2.66$\pm$0.52\\
      47.2&      5.21$\pm$0.81&      4.92$\pm$0.71&      3.39$\pm$0.59&      3.80$\pm$0.63&      2.39$\pm$0.51&      1.05$\pm$0.32\\
      54.6&      6.63$\pm$0.91&      5.45$\pm$0.75&      3.43$\pm$0.59&      3.77$\pm$0.63&      2.97$\pm$0.57&      3.23$\pm$0.57\\
      63.9&      9.70$\pm$1.13&      8.72$\pm$0.96&      6.98$\pm$0.86&      3.30$\pm$0.60&      6.75$\pm$0.88&      5.29$\pm$0.74\\
      76.0&      8.39$\pm$1.09&      9.82$\pm$1.06&      6.66$\pm$0.87&      4.90$\pm$0.76&      4.00$\pm$0.70&      3.32$\pm$0.61\\
	\bottomrule	
	
\end{longtable}
\section{Appendix B}
\begin{figure}
	\centering
	\includegraphics[scale=0.7]{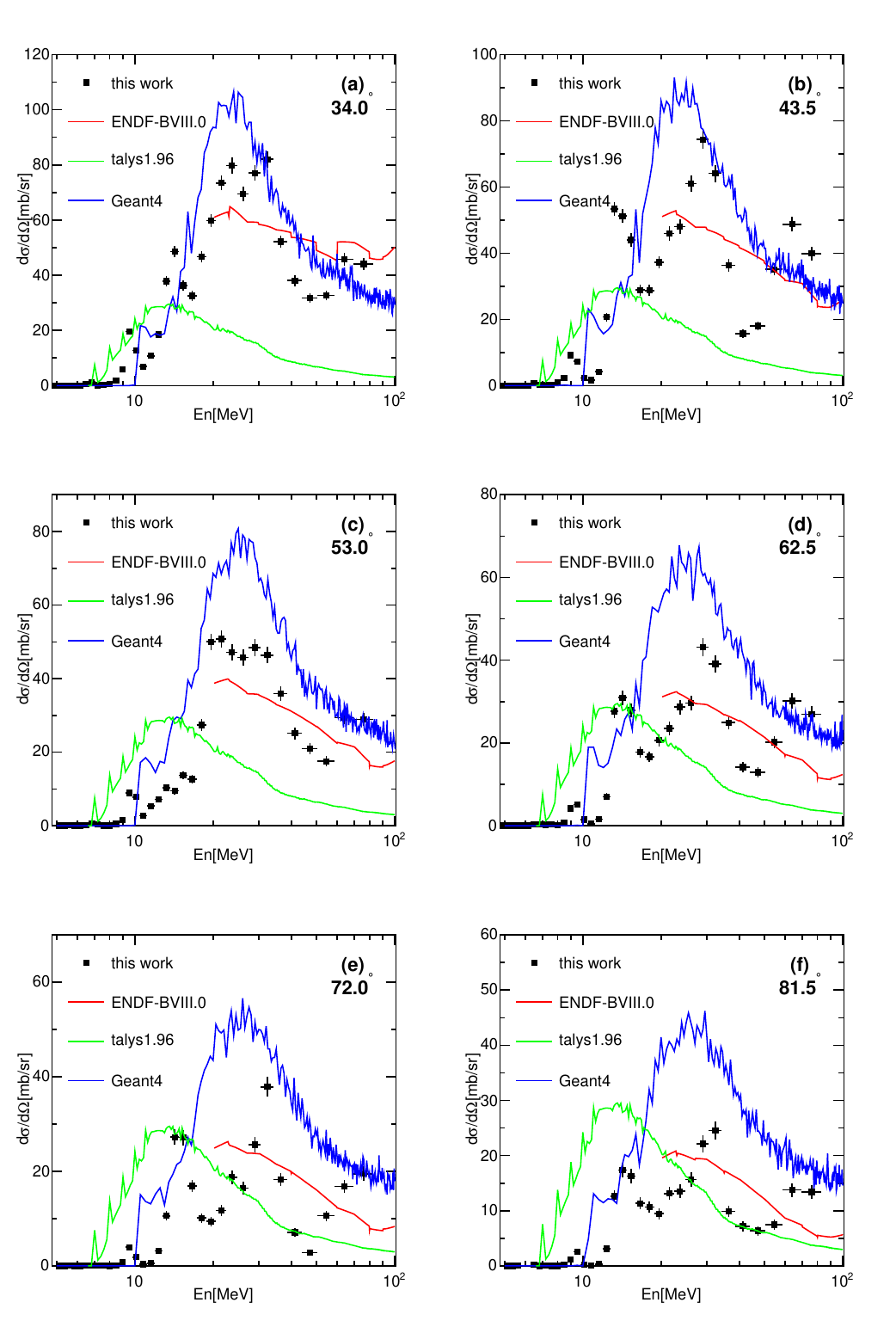}
	\caption{\centering (color online) Differential cross sections of the $^{12}C(n,\alpha)x$ reactions measured at $\theta_{L} = 34.0^{\circ}$,$\theta_{L} = 43.5^{\circ}$,$\theta_{L} = 53.0^{\circ}$,$\theta_{L} = 62.5^{\circ}$,$\theta_{L} = 72.0^{\circ}$,$\theta_{L} = 81.5 ^{\circ}$ compared with results from the evaluated data, and the simulation.(Only statistical uncertainties were included)}
	\label{aMACS1_p}
\end{figure}
\begin{figure}
	\centering
	\includegraphics[scale=0.7]{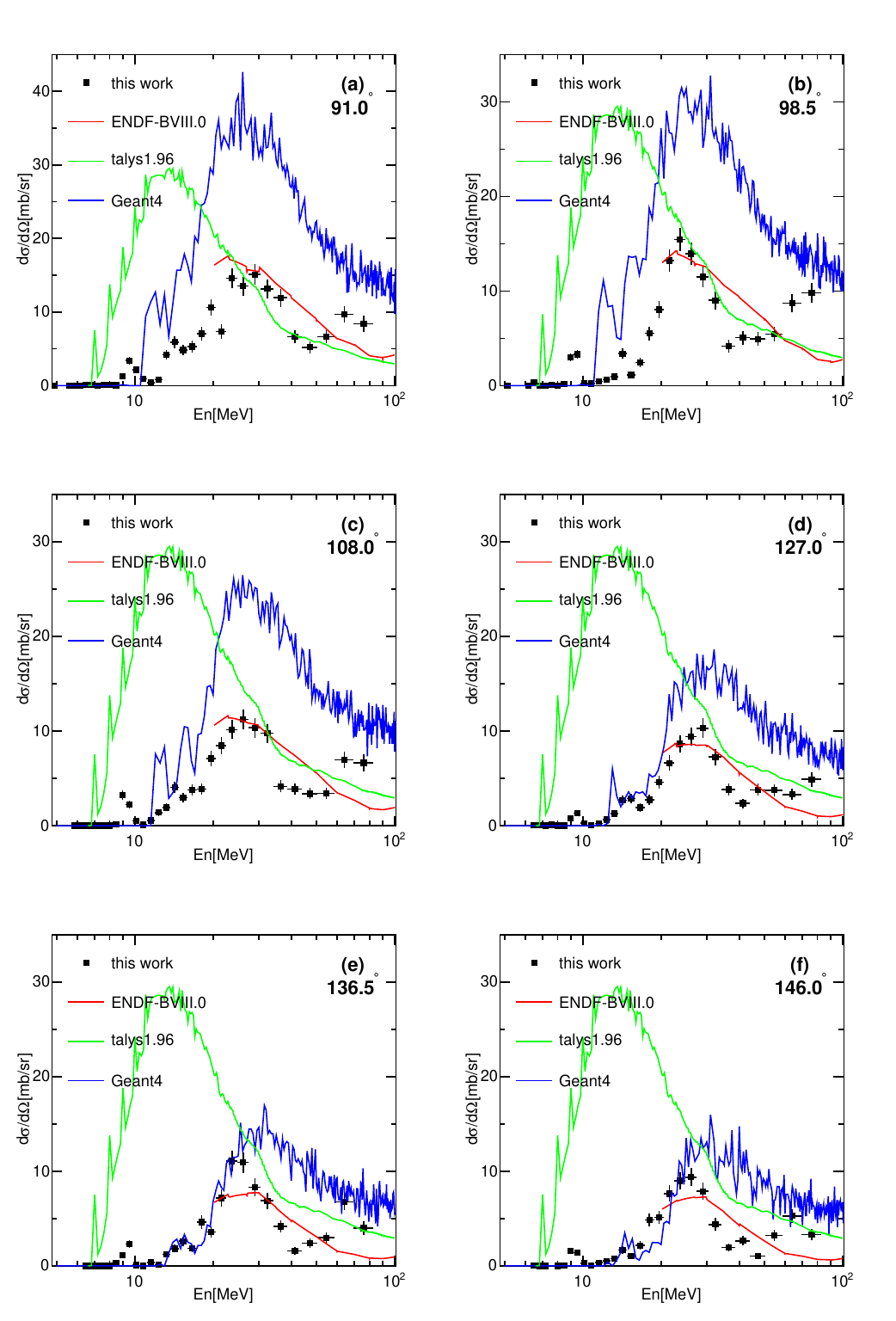}
	\caption{\centering (color online) Differential cross sections of the $^{12}C(n,\alpha)x$ reactions measured at $\theta_{L} = 91.0^{\circ}$,$\theta_{R} = 98.5^{\circ}$,$\theta_{R} = 108.0^{\circ}$,$\theta_{R} = 127.0^{\circ}$,$\theta_{R} = 136.5^{\circ}$,$\theta_{R} = 146.0 ^{\circ}$compared with results from the evaluated data, and the simulation.(Only statistical uncertainties were included)}
	\label{aMACS2_p}
\end{figure}

\end{appendices}
\vskip3pt
\end{document}